\documentclass{IEEEtran}
\usepackage{cite}
\usepackage{amsmath,amssymb,amsfonts}
\usepackage{algorithmic}
\usepackage{graphicx}
\usepackage{subcaption}
\usepackage{textcomp}
\usepackage{generic}
\usepackage{multirow}

\usepackage{tikz}
\usepackage{lipsum}

\def\BibTeX{{\rm B\kern-.05em{\sc i\kern-.025em b}\kern-.08em
    T\kern-.1667em\lower.7ex\hbox{E}\kern-.125emX}}
\begin{document}
\title{Digital-Based Potentiostat and Mesoporous Microelectrode Co-Design for Non-Enzymatic Glucose Detection at 0.3\,V-$V_{\mathrm{DD}}$  and 1.65\,nW-Power}
\author{Andrea De Gregorio, Mara Serrapede, Danilo Kaddouri, Paolo Angelini, Giuseppe Bruno, Simone Luigi Marasso, Salvatore Guastella, Andrea Lamberti, and Paolo Crovetti, \IEEEmembership{Senior Member, IEEE}
\thanks{
Andrea De Gregorio and Paolo Crovetti are with the Department of
Electronics and Telecommunication (DET), Politecnico di Torino, 10129 Turin, Italy (emails: andrea.degregorio@polito.it, paolo.crovetti@polito.it)\\
Mara Serrapede, Simone Luigi Marasso, Salvatore Guastella and Andrea Lamberti are with the Departement of Applied Science and Technology (DISAT), Politecnico di Torino, 10129 Turin, Italy (emails: mara.serrapede@polito.it, simone.marasso@polito.it, salvatore.guastella@polito.it, andrea.lamberti@polito.it)\\
Danilo Kaddouri, Paolo Angelini and Giuseppe Bruno are with STMicroelectronics, Milan, Italy (email: danilo.kaddouri@st.com, paolo.angelini@st.com, giuseppe.bruno@st.com)\\
This publication is part of the project PNRR-NGEU which has received funding from MUR - DM 117/2023. 
This study was carried out within the MicroBioNIC project within the PRIN 2022 program (D.D. 104 - 02/02/2022 Ministero dell’Università e della Ricerca). This manuscript reflects only the authors’ views and opinions and the Ministry cannot be considered responsible for them. 
This work was partially funded under the National Recovery and Resilience Plan (NRRP), Mission 4 “Education and Research”-Component 2 “From research to business”- Investment 3.1 “Fund for the realization of an integrated system of research and innovation infrastructures”- Call for tender No. N. 3264 of 28/12/2021 of Italian Ministry of Research funded by the European Union-NextGenerationEU Project code: IR0000027, Concession Decree No. 128 of 21/06/2022 adopted by the Italian Ministry of Research, CUP: B33C22000710006, Project title: iENTRANCE.}}

\maketitle

\begin{abstract}
This paper presents a proof-of-concept ultra-low voltage and ultra-low power chronoamperometric electrochemical sensor for non-enzymatic glucose readout integrated circuit (IC) in 130\,nm CMOS  detection featuring a reconfigurable Digital-Based (DB) Potentiostat. The signal transfer and noise characteristics of the new digital-based architecture are analytically described in the frequency domain for the first time by an equivalent linearized model that is validated by simulations and experiments. 
Based on experiments, the proposed DB potentiostat enables the detection of a wide electrochemical current range, spanning from 600\,pA to 650\,nA, with $R^2$=0.991 linearity and consumes only 1.65\,nW (53.5\,nW) at $V$\textsubscript{dd} = 300\,mV ($V$\textsubscript{dd} = 500\,mV). The proposed DB readout is tested in a proof-of-concept platform for non-enzymatic glucose detection with nanostructured microelectrodes, demonstrating successful non-enzymatic glucose detection at physiological levels at the lowest reported voltage and power, even in the presence of an interferent (ascorbic acid) and under aerobic conditions, thus revealing a strong potential for emerging Point of Care (PoC) diagnostics applications.
\end{abstract}


\newcommand\copyrighttext{%
  \footnotesize This work has been submitted to the IEEE for possible publication. Copyright may be transferred without notice, after which this version may no longer be accessible.}
\newcommand\copyrightnotice{%
\begin{tikzpicture}[remember picture,overlay]
\node[anchor=south,yshift=10pt] at (current page.south) {\fbox{\parbox{\dimexpr\textwidth-\fboxsep-\fboxrule\relax}{\copyrighttext}}};
\end{tikzpicture}%
}
\copyrightnotice

\newpage
\section{Introduction}

\begin{figure}[t]  
    \centering
    \includegraphics[width=0.5\textwidth]{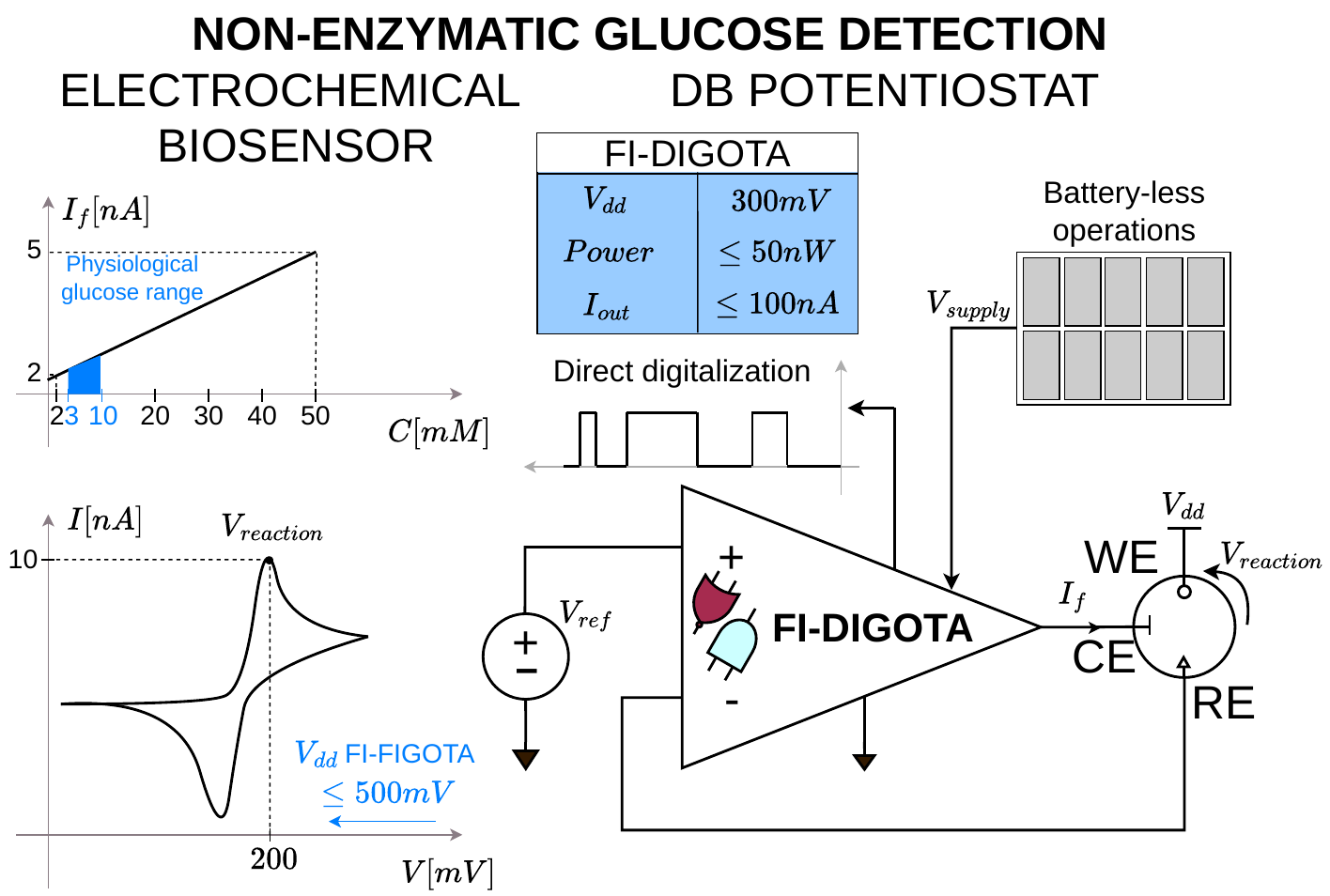} 
    \caption{Representation of electrochemical biosensors constraints for non-enzymatic glucose detection, satisfied by DB Potentiostat features.}
   \label{Fig1_intro}
\end{figure}

\IEEEPARstart{E}{merging} wearables, personalized healthcare and Point-of-Care (PoC) diagnostics require robust, untethered, energy-autonomous electrochemical sensing platforms capable of delivering reliable real-time information on the concentration of clinically relevant analytes such as glucose, lactate and drugs \cite{1_1_Kim2019}\cite{3_26_QU}\cite{4_27_AGUILERAVENEGAS20125837}\cite{5_28_Kocak_2023}\cite{6_29_WANG2023124362}.


Meeting such requirements in electrochemical sensors results in many challenges both in the design of the sensing electrodes and in the readout integrated circuits (ICs), that need to be addressed in a synergistic way  \cite{2_48_Carrara2013}.

On one side, conventional electrochemical sensors that take advantage of enzyme-mediated reactions to enhance selectivity are not attractive due to their stringent limitations in operating and storage  conditions (e.g. temperature, humidity, presence of chemicals), potential enzyme instability, dependence on cofactors and stabilizers and susceptibility to interferents  \cite{7_30_LEE201824}\cite{8_31_Niyazi2025}\cite{9_32_Kim2022}. On the other side, focusing on the readout, energy autonomous miniaturized sensing platforms demand integrated circuits (ICs) operating at Ultra-Low Voltage (ULV) and Ultra-Low Power (ULP) consumption (up to tens of nW) to be supplied by an unregulated energy harvester or to maximize the lifetime of miniaturized batteries \cite{18_2_Alioto2017} \cite{19_3_Alioto2022}.

Aiming to overcome the limitations of enzymatic sensing, non-enzymatic electrochemical biosensors enabling direct oxidation or reduction of the analyte at the electrode surface have been proposed and investigated from decades \cite{3_26_QU}\cite{4_27_AGUILERAVENEGAS20125837}\cite{5_28_Kocak_2023}\cite{6_29_WANG2023124362}\cite{8_31_Niyazi2025}\cite{9_32_Kim2022}\cite{10_33_DONG2023120463}. 
Among these materials, platinum (Pt) and Pt-based composites are widely used due to their excellent electrocatalytic activity toward small biomolecules, high conductivity, chemical stability under physiological conditions, and general biocompatibility \cite{5_28_Kocak_2023}\cite{6_29_WANG2023124362}\cite{7_30_LEE201824}\cite{8_31_Niyazi2025}\cite{9_32_Kim2022}. Consequently, significant efforts have focused on engineering Pt nanostructures and surface chemistries to boost catalytic efficiency, lower overpotentials, and enhance resistance to interferents \cite{3_26_QU}\cite{4_27_AGUILERAVENEGAS20125837}\cite{5_28_Kocak_2023}\cite{6_29_WANG2023124362}\cite{8_31_Niyazi2025}\cite{9_32_Kim2022}\cite{10_33_DONG2023120463}. 



From the IC readout point of view, conventional potentiostats for amperometric readout based on current-to-voltage conversion by  resistive \cite{21_5_alMamun}\cite{22_6_Lin}\cite{23_7_Djekic}  or capacitive \cite{24_8_Chilukuri}\cite{25_9_Ghoreishizadeh}\cite{26_10_Hsu}\cite{27_11_Wang}\cite{28_12_valente} feedback transimpedance amplifiers (TIA) and analog-to-digital converters (ADCs) often require impractically large passives and/or low noise, high performance amplifiers which are not compatible with low voltage and low power operation. Alternatives based on current conveyors (CCs) \cite{29_13_Ahmad}\cite{30_14_Li}\cite{31_15_Dorta}\cite{32_16_Kang} extend the input range at the cost of a further increased power consumption, and solutions based on current-to-frequency conversion \cite{33_17_Yu}\cite{34_18_Lin}\cite{35_19_chou}, enable the direct digitization of the sensed currents without external ADCs but often require high-speed comparators and counters resulting in marginal power savings and increased complexity.

More recently, wide dynamic range and nW-power operation have been demonstrated in digital-based (DB) potentiostats \cite{36_20_rubino}\cite{37_21_akram}. In details, DB potentiostats \cite{36_20_rubino} \cite{40_24_DeGregorio} \cite{41_25_Meimandi} inherit the advantages of the Digital Operational Transconductance Amplifier (DIGOTA) \cite{38_22_crovetti}\cite{39_23_TOledo} and of the Floating Inverter (FI)-based Digital Operational Transconductance Amplifier (OTA) (FI-DIGOTA) \cite{36_20_rubino}\cite{40_24_DeGregorio} and achieve both ULV and ULP operations, thanks to bias current suppression, near-threshold dynamic operations, and direct digitization without external ADCs.



The electrochemical sensors proposed for non-enzymatic glucose monitoring so far, however, require reference voltages well above the minimum power supply voltage of a DB-potentiostat (300\,mV), and the faradaic current drawn by standard centimeter-sized electrodes is well above the tens of microampere range, thus keeping the minimum voltage requirements above 1\,V and the power at the tens of microwatts and  hampering the inherent advantages of the digital-based approach.

In this paper, a proof-of-concept ULV/ULP sensing platform for non-enzymatic glucose detection based on the co-design of a DB-potentiostat and a new  nanostructured mesoporous platinum microelectrode fabricated according to the method firstly proposed by Attard et al. \cite{13_36_Attard} is proposed. The adoption of such specific nanostructured microelectrodes and digital-based readout is shown to enable non-enzymatic glucose detection at physiological levels with nA-range faradaic current and a reference voltage in the sub-300\,mV range, in line with emerging biosensor requirements.

The rest of the paper is organized as follows: in Sect. II the architecture of the FI-DIGOTA is briefly revisited and in Section III the signal transfer and noise characteristics of a DB-potentiostat are analytically described for the first time by an equivalent frequency-domain linearized model that is validated by simulations and experiments.


The design and fabrication of the new mesoporous microelectrodes are described in Sect. IV while, in Sect. V the electrochemical readout capabilities of the DB Potentiostat are validated through the sensing of the ferrocyanide oxidation current. In Sect. VI the operation of the overall proof-of-concept sensing platform for non-enzymatic glucose sensing is experimentally demonstrated and, finally, in Section VII, some concluding remarks are drawn.

\section{The FI-DIGOTA Circuit}

\begin{figure}[t]  
    \centering
    \includegraphics[width=0.5\textwidth]{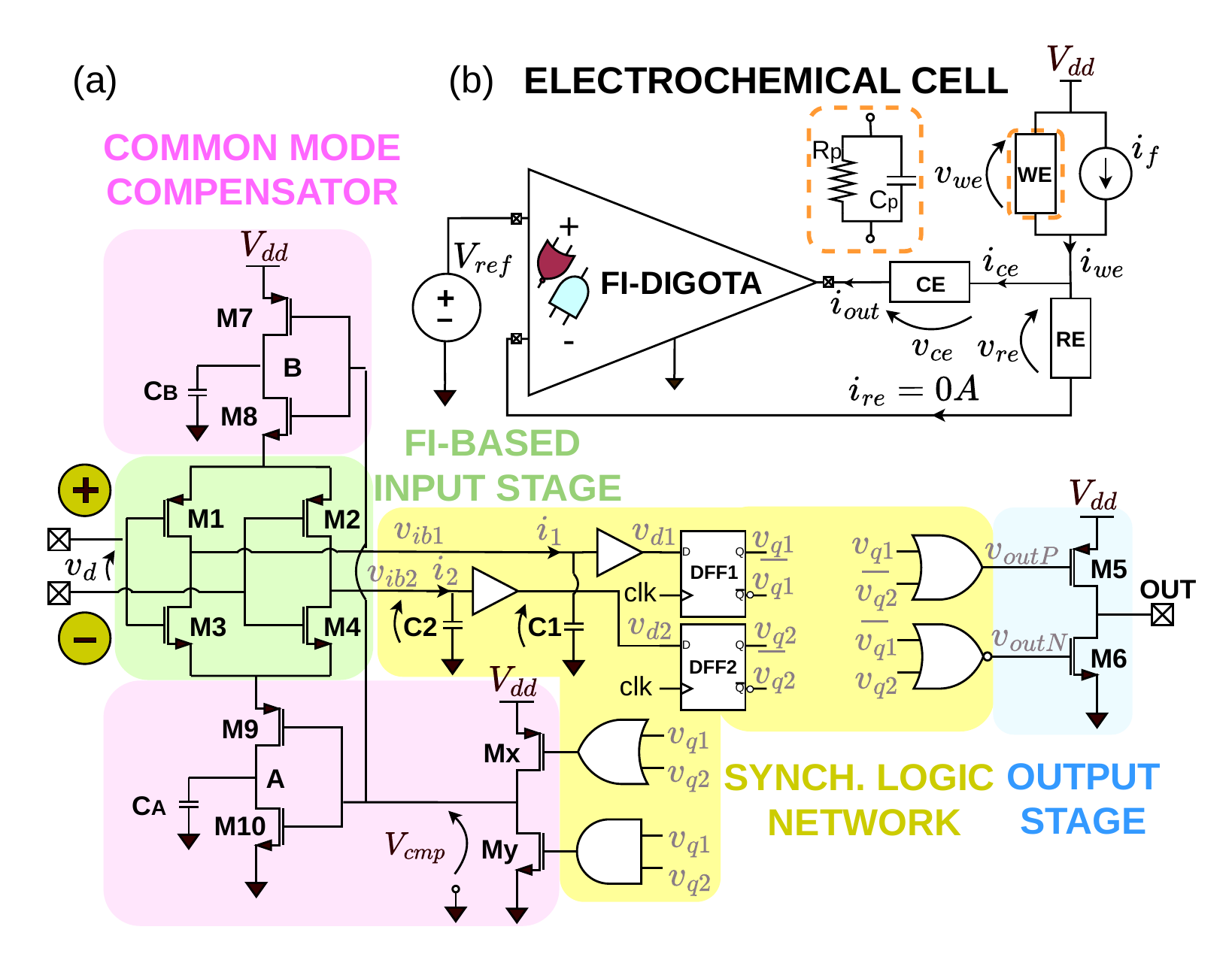} 
    \caption{(a) FI-DIGOTA circuit schematic, (b) DB Potentiostat: FI-DIGOTA in feedback with an electrochemical cell.}
   \label{Fig2_FIDIGOTA_Schematic}
\end{figure}

To enable non-enzymatic glucose monitoring in wearable and PoC diagnostic applications, the DB-potentiostat in Fig.\ref{Fig2_FIDIGOTA_Schematic} is selected for its nanowatt-level power consumption and its operation at supply voltages below 500\,mV. The architecture is based on the topology originally proposed in \cite{36_20_rubino} and later adapted  for dopamine sensing in \cite{41_25_Meimandi}.


In this section, the operation of the FI-DIGOTA is reviewed, and its small-signal frequency-domain equivalent model is derived following the approach in \cite{39_23_TOledo}. This model is then used in Sect.\,III to derive the signal and noise transfer functions of the DB potentiostat.

\begin{figure}[t]  
    \centering
    \includegraphics[width=0.48\textwidth]{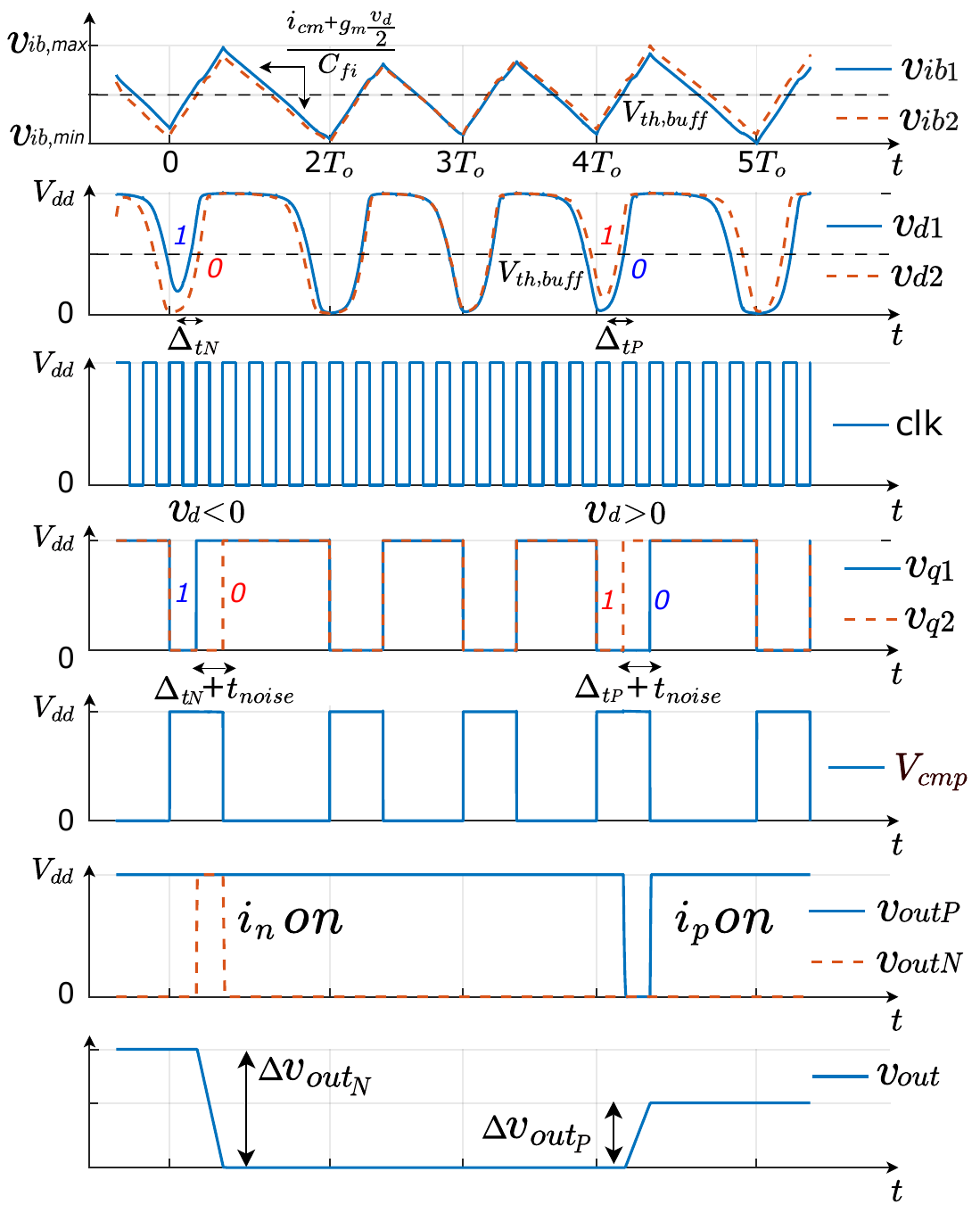} 
    \caption{FI-DIGOTA internal waveforms.}
   \label{Fig3_waveforms}
\end{figure}


\subsection{Architecture and Operating Principle}
The FI-DIGOTA in Fig.\ref{Fig2_FIDIGOTA_Schematic} comprises a Floating Inverter (FI) based input stage (M1 – M4),
a synchronous logic network, a Common Mode (CM) compensation network (Mx, My, M7 – M10) and a three-state output stage (M5 – M6). In the proposed implementation, the strength of the nMOS and pMOS devices of the output stage can be independently configured by two 8-bit digital calibration words \cite{40_24_DeGregorio}.


The behavior of the internal waveforms of the circuit is depicted in Fig.\ref{Fig3_waveforms}.
The outputs of the FI-based input stage ($v$\textsubscript{ib1}, $v$\textsubscript{ib2}) are converted into two-level signals by comparison with the threshold voltages $V_{th,buff}$ of two digital buffers. These signals are then sampled by two D-type flip-flops, resulting in four possible states for ($v$\textsubscript{q1}, $v$\textsubscript{q2}): (0,0), (0,1), (1,0), and (1,1).



According to the algorithm described in \cite{38_22_crovetti}\cite{39_23_TOledo}, when $v$\textsubscript{q1} and $v$\textsubscript{q2} assume the same logical value, the sign of the differential input voltage $v$\textsubscript{d} cannot be inferred and the CM compensation network is activated to provide dynamic biasing to the FI input stage. If $v$\textsubscript{q1} = $v$\textsubscript{q2} = 0 ($v$\textsubscript{q1} =  $v$\textsubscript{q2} = 1) the positive (negative) supply of the input stage is connected through M8 (M9) to the capacitance Cb (Ca), that was pre-charged at $V\textsubscript{dd}$ (0V) via M7 (M10). In the absence of a differential input component, the FI-DIGOTA operates in a self-oscillating mode, periodically switching between (0,0) and (1,1) states, with a self-oscillation period ($T$\textsubscript{0}) dependent on the charge/discharge rate of the parasitic output capacitors (C1, C2) of the FI, as detailed in \cite{39_23_TOledo}.

When a differential input is applied, it can be modeled as a perturbation superimposed on the steady-state oscillation. This perturbation gives rise to opposite variations in the charging slopes $v$\textsubscript{ib1} and $v$\textsubscript{ib2}, resulting in a timing difference ($\Delta t$ in Fig\ref{Fig4_BlockDiagram}(a)) between the switching events of the following digital buffers that is proportional to the input differential voltage $v$\textsubscript{d}. The time difference between the transitions of digital signals $v$\textsubscript{d1} and $v$\textsubscript{d2} is then processed by the synchronous logic network that turns on the output stage accordingly.


With respect to \cite{39_23_TOledo}, to obtain a direct digitalization of the output through the digital streams $v$\textsubscript{outP} and $v$\textsubscript{outN} in Fig.\ref{Fig2_FIDIGOTA_Schematic}, the proposed architecture presents two D flip-flops (D-FFs) that oversample the signals $v$\textsubscript{d1} and $v$\textsubscript{d2}, resulting in a quantization of $\Delta t$ into integer multiples of the clock period. The outputs of the D-FFs ($v$\textsubscript{q1}, $v$\textsubscript{q2}) drive the rest of the circuit according to the sign of the differential input voltage $v$\textsubscript{d}.
In particular, if $v$\textsubscript{q1} = 0 and $v$\textsubscript{q2} = 1 ($v$\textsubscript{q1} = 1 and $v$\textsubscript{q2} = 0), $v$\textsubscript{d} is positive (negative) and the pMOS M5 (nMOS M6) is turned on to sink (source) current to (from) the load.

\subsection{FI-DIGOTA Frequency Response}

\begin{figure*}[t]  
    \centering
    \includegraphics[width=1\textwidth]{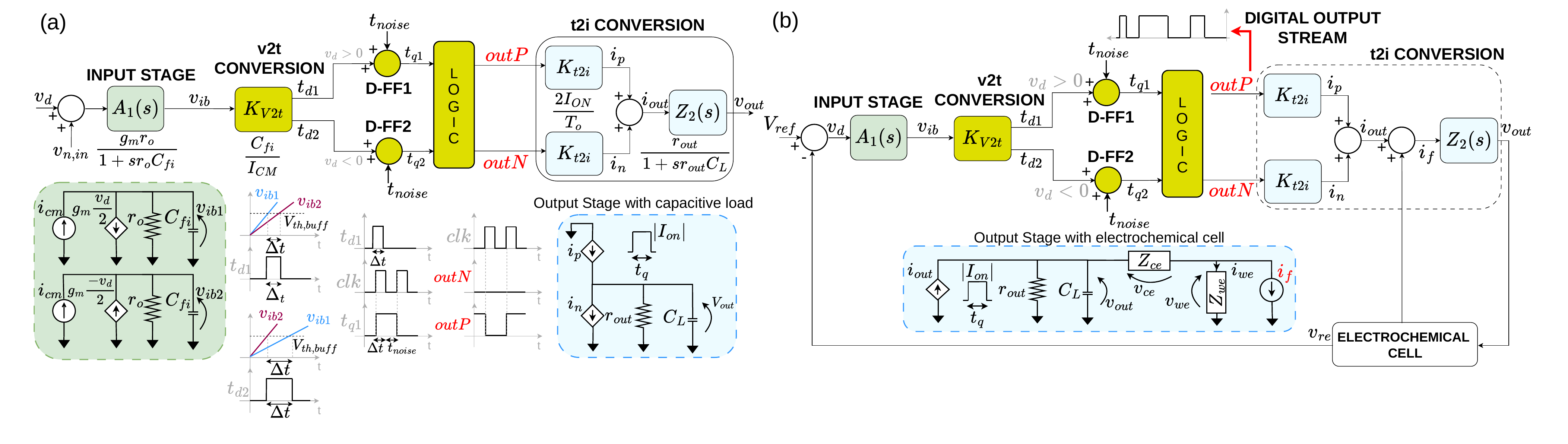} 
    \caption{(a) FI-DIGOTA open loop block diagram (b) DB Potentiostat block diagram.}
   \label{Fig4_BlockDiagram}
\end{figure*}

The frequency-domain analysis of the FI-DIGOTA is performed adapting the modeling approach in \cite{39_23_TOledo}, in consideration of the two main differences among the DIGOTA circuit considered in \cite{39_23_TOledo} and the FI-DIGOTA considered in this work, namely: the different input stage (Muller-C input stage in \cite{39_23_TOledo} vs. FI-based input in this work), and the synchronous operation of the FI-DIGOTA, as opposed to the asynchronous operation of the DIGOTA in \cite{39_23_TOledo}.

Under the same assumptions in \cite{39_23_TOledo}, both DIGOTA and FI-DIGOTA can be described as the cascade connection of an equivalent transconductance input stage followed by a voltage-to-time conversion block, which converts the input differential voltage into a proportional timing difference $\Delta t$ and by a time-to-current conversion block, that converts $\Delta t$ into current pulses driving an $RC$ output impedance, as shown in the block diagram in Fig.\ref{Fig4_BlockDiagram}(a).






Although the Muller-C input stage in \cite{39_23_TOledo} and the FI- based input stage considered in this work are different, under small-signal conditions, the perturbation in their output currents related to the input differential voltage can be described in both cases in terms of a time-varying transconductance, whose averaged value is fairly constant over each CM-compensation cycle, that drives an equivalent $RC$ load representing the output resistance of the FI stage $r_o$ in parallel to the parasitic capacitance at the FI output node, $C_{fi}$ resulting in a one-pole transfer function.


As detailed for the FI-DIGOTA operating principle, the effect of $v$\textsubscript{d} determines a change in the slopes $\frac{I_{cm}\pm g_m v_d/2}{C_{\mathrm{{fi}}}} $ of $v$\textsubscript{ib1} and $v$\textsubscript{ib2} signals, leading to a time difference ($\Delta t$) in the point of time in which these signals cross the threshold voltages of the two digital buffers ($V$\textsubscript{th,buff}).
Thereby, the time difference between the commutations of $v$\textsubscript{d1} and $v$\textsubscript{d2} signals is directly proportional to $\Delta t$ and, in the block diagram in Fig.\ref{Fig4_BlockDiagram}(a), it is represented by the $t$\textsubscript{d} signals that are obtained, from an analytical point of view, multiplying the input stage transfer function times the ratio $C$\textsubscript{fi}/$I$\textsubscript{cm}, where $I$\textsubscript{cm} is the  common mode current injected in $C$\textsubscript{fi} during the states (0,0),(1,1).
This mechanism, depicted in Fig.\ref{Fig4_BlockDiagram}(a), brings the two digital buffers to perform a voltage-to-time (V2t) conversion.


Unlike in the DIGOTA considered in \cite{39_23_TOledo}, where the signals which contain the information about $v$\textsubscript{d} drive the output stage directly, in the FI-DIGOTA  $v$\textsubscript{d1} and  $v$\textsubscript{d2} are sampled by the D-FFs, so that their transitions are synchronized with the nearest clock edge, resulting in the  $v$\textsubscript{q} signals in which the duration of the time intervals between two transitions of $v$\textsubscript{ib1} and $v$\textsubscript{ib2}, which conveys the information about $v$\textsubscript{d}, is constrained to be an integer multiple of the clock cycle. The time-domain information is therefore affected by a quantization error, that is described in our model as an additive white noise ($t$\textsubscript{noise}) with uniform amplitude distribution in the (0, 2$T$\textsubscript{clk}) range, uncorrelated to the input signal, under the same assumptions routinely considered in the analysis of oversampled data converters \cite{DeLaRosa_sigmaDelta}. 



As in \cite{39_23_TOledo}, the output stage, when active, sources (sinks) a nearly constant current $i$\textsubscript{p} ($i$\textsubscript{n}) into the output impedance $Z_2$ during $t_{q1}$ ($t_{q2}$), thus effectively performing a time-to-current (t2i) conversion and, overall, translating the input differential voltage ($v$\textsubscript{d}) into a current. Following the same approach in \cite{39_23_TOledo} the output stage can be described in the frequency domain as a parallel $RC$ impedance driven by a pulsed current and contributes with a pole to the FI-DIGOTA open-loop gain.



\section{DB Potentiostat Operation and Frequency-Domain Analysis}

This section describes the operation of the DB potentiostat and derives its frequency-domain model based on the equivalent representation of the electrochemical cell.

\subsection{Electrochemical Cell Model}
A standard three-electrode configuration (WE, CE, RE) is assumed. The WE is modeled using a Randles equivalent extracted from the characterization of the electrode introduced in Section III-A. The charge-transfer resistance is in parallel with a Constant Phase Element (CPE), as shown in Fig.\ref{Fig2_FIDIGOTA_Schematic}. The solution resistance, which is in the k$\Omega$ range, is neglected in consideration of the negligible voltage drop across it for nA-range currents \cite{42_x_Franks}.

Experimental characterization of the nanostructured microelectrode (Section IV) yields $C$\textsubscript{p} = 7\,nF and $R$\textsubscript{p} = 220\,M$\Omega$, consistent with reported values \cite{42_x_Franks,43_x_Elliott}. The CE impedance is assumed negligible with respect to the WE and is therefore omitted.

\subsection{DB Potentiostat Static and Dynamic Analysis}

The proposed DB potentiostat (Fig.\ref{Fig2_FIDIGOTA_Schematic}(b)) embeds the FI-DIGOTA in a feedback loop with the electrochemical cell for amperometric sensing.

The voltage at the reference electrode (RE), $v_{\text{re}} = v_{-}$, is regulated to $V_{\text{ref}} = v_{+}$ by sourcing or sinking discrete charge packets $i_{p}T_{\text{clk}}$ and $i_{n}T_{\text{clk}}$ at the counter electrode (CE), where $i_{p}$ and $i_{n}$ are the output-stage currents.

Under static conditions, the average output current equals the faradaic current $i_f$, which can be expressed considering the number of active digital pulses (p,n) over an acquisition window of $M T_{\text{clk}}$ as
\begin{equation}
\label{eq1_FaradicCurrent}
 i_f = \frac{p i_{p} - n i_{n}}{M}.
\end{equation}

While (\ref{eq1_FaradicCurrent}) has been used in prior work \cite{36_20_rubino}, \cite{41_25_Meimandi}, a frequency-domain signal and noise analysis of the DB potentiostat is still lacking and is developed in the following.

\subsection{DB Potentiostat Frequency-Domain Analysis}

Based on the small-signal model of the FI-DIGOTA in Fig.~\ref{Fig4_BlockDiagram}(a) and on the Randles equivalent of the electrochemical cell, the frequency-domain behavior of the DB potentiostat can be derived.

Referring to the block diagram in Fig.~\ref{Fig4_BlockDiagram}(b), the quantized pulse width $t_q$ conveys the information associated with the timing difference $\Delta t$, while the effect of time quantization is modeled as an additive noise contribution, which is taken into account along with the input-referred noise source $v_{n,in}$ of the potentiostat.

The characterization of the DB potentiostat is carried out by analyzing the variation of $t_q$ as a function of the faradaic current $i_f$ for the signal transfer function, and by considering the contribution of the quantization noise ($t_{\text{noise}}$) and of the FI-DIGOTA input-referred noise  ($v_{\text{n,in}}$) for the noise transfer functions. For simplicity, and consistently with the considered operating conditions ($i_f > 0$), only the pull-up transistor M5 is included in the output-stage model.




\subsubsection{Signal Transfer Function}

From the block diagram in Fig.~\ref{Fig4_BlockDiagram}(b), the signal transfer function relating the digital pulse width $t_{q1}$ to the faradaic current ($i_f = i_{\text{we}} = i_{\text{ce}}$) can be expressed as

\begin{eqnarray}
\label{eq3_signalTF}
 \frac{t_{q1}}{i_f} &=&  \frac{Z_2(s) A_1(s) k_{v2t} }{1+A_1(s) k_{v2t} k_{t2i} Z_2(s)}\\
 &=&\frac{g_m r_o \frac{C_{fi}}{I_{cm}} r_{out}}{(1 + s r_o C_{fi})(1 + s r_{out} C_L)+ 2 g_m r_o \frac{C_{fi}}{I_{cm}}r_{out} \frac{I_{on}}{T_0}} \nonumber.
\end{eqnarray}

The DB potentiostat can be modeled as a second-order system. The low-frequency gain is determined by the intrinsic gain of the input stage, the ratio $C_{fi}/I_{cm}$, and the output resistance $r_{out}$, while the pole frequencies are set by the time constants $r_o C_{fi}$ and $r_{out} C_L$.





\begin{figure}[t]  
    \centering
    \includegraphics[width=0.5\textwidth]{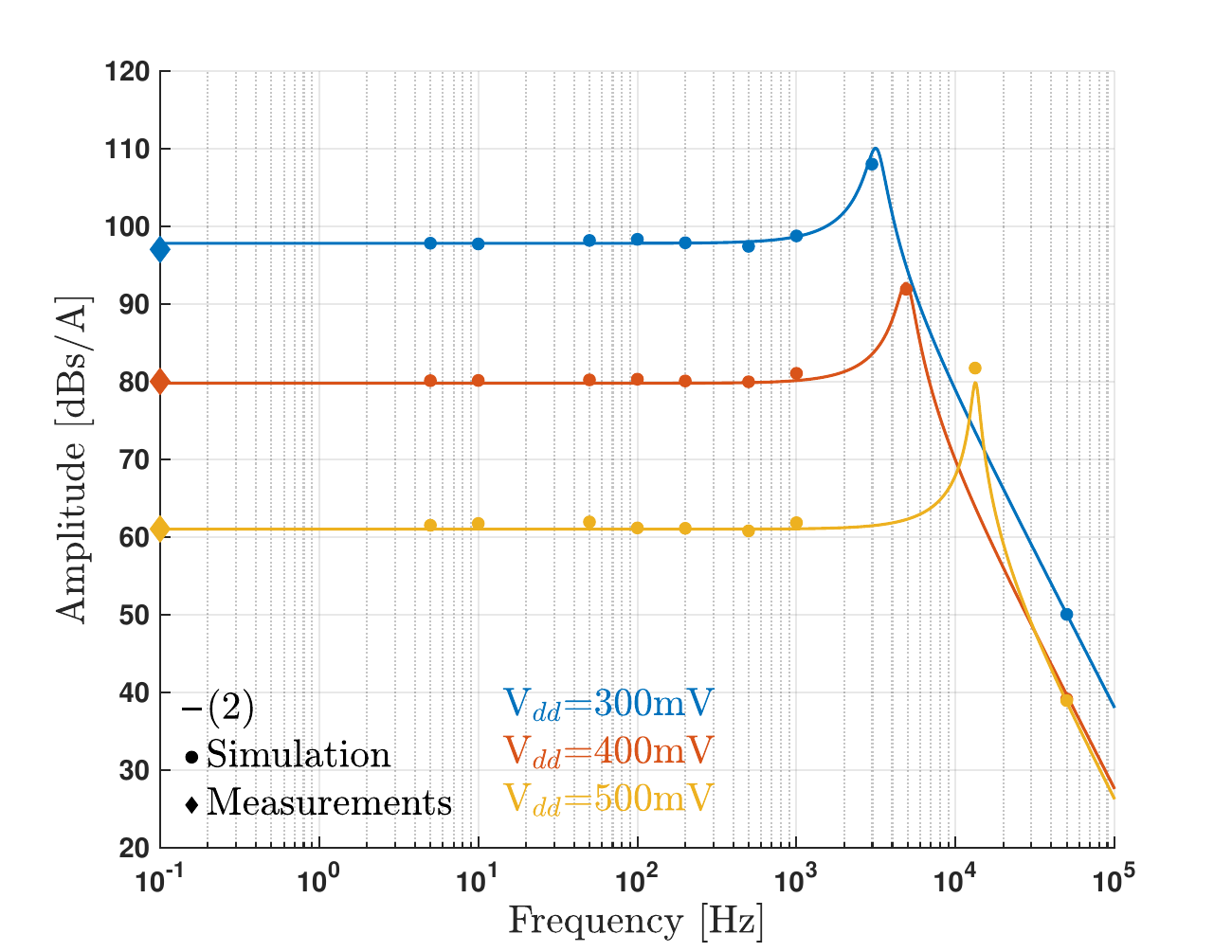} 
    \caption{DB Potentiostat signal transfer function with different $V$\textsubscript{dd}, comparison between analytical model, simulations and measurements.}
   \label{Fig5_SignalTF}
\end{figure}

Fig.\ref{Fig5_SignalTF} shows the comparison of the model described in (\ref{eq3_signalTF}) with simulations and measurements.

\begin{table}
\centering
\caption{FI-DIGOTA Simulations Parameters ($V$\textsubscript{dd} = 0.4\,V)}
\label{table2}
\setlength{\tabcolsep}{3pt}
\begin{tabular}{|c|c|c|}
\hline
Parameter &  Unit &Value \\ \hline
 $T$\textsubscript{0} & $\mu$s & 103  \\ \hline
 $g$\textsubscript{m} & nS  & 61 \\ \hline
 $r$\textsubscript{0} & G$\Omega$ &  89 \\ \hline
 $r$\textsubscript{out} & k$\Omega$ &  102 \\ \hline
 $C$\textsubscript{fi} & fF &  1.9 \\ \hline
 $I$\textsubscript{on} & nA &  8.1 \\ \hline
 $I$\textsubscript{cm} & pA &  0.8 \\ \hline
\end{tabular}
\end{table}

The parameters used to evaluate (\ref{eq3_signalTF}) are reported in Table~\ref{table2} for $V$\textsubscript{dd} = 0.4\,V. The frequency behavior of $t_{q1}$ is obtained from the FFT of the normalized $v$\textsubscript{outP} signal. Simulations are performed by sweeping $V$\textsubscript{dd} while maintaining a fixed clock frequency of 50\,kHz. The amplitude of $i_f$ and the strength of the output stage are adjusted accordingly.

Since the DB potentiostat operates in chronoamperometric mode, the measured $v$\textsubscript{outP} spectrum reflects the DC response of (\ref{eq3_signalTF}). The FFT of the $v$\textsubscript{outP} signals obtained from ferrocyanide measurements (Section V) is reported in Fig.\ref{Fig5_SignalTF} and compared with simulations and the analytical model.

As $V$\textsubscript{dd} decreases, the DC gain increases, resulting in higher sensitivity of the digital output to variations of the faradaic current. In the subthreshold regime, this behavior arises from the exponential dependence of the average output current $I$\textsubscript{on}, to which the transfer function is inversely related.
Since the second pole is only weakly dependent on $V$\textsubscript{dd}, the overall gain-bandwidth product slightly improves at lower supply voltages.
Fig.\ref{Fig5_SignalTF} shows good agreement between the model, simulations, and measurements over the entire frequency range.


\subsubsection{Noise Transfer Function}

\begin{figure}[t]  
    \centering
    \includegraphics[width=0.5\textwidth]{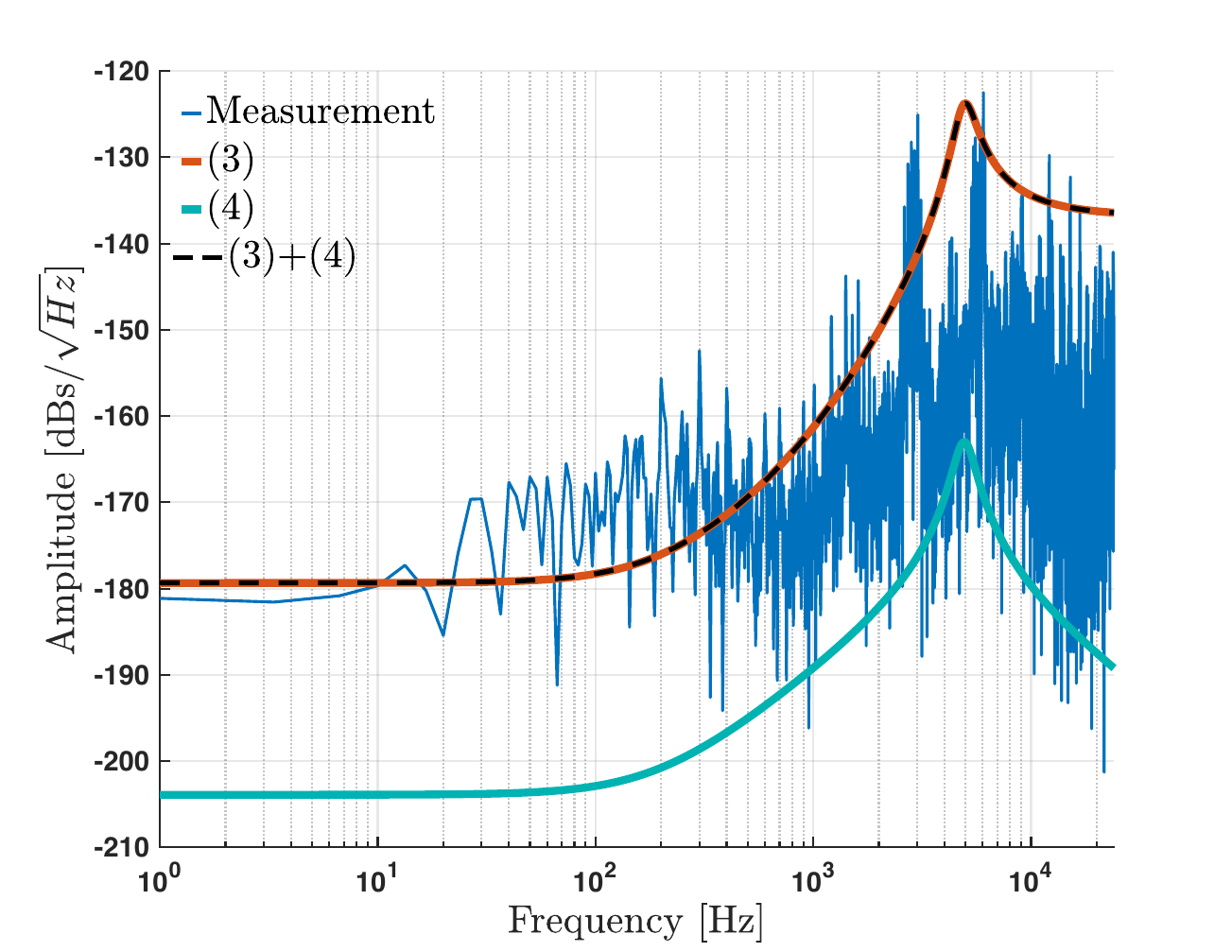} 
    \caption{DB Potentiostat noise power spectrum at $V$\textsubscript{dd} = 0.4\,V, comparison between analytical model and measurement.}
   \label{Fig6_NoiseTF}
\end{figure}

Considering Fig.\ref{Fig4_BlockDiagram}(b), the noise transfer function (NTF) relating the quantization noise introduced by the by the D-FFs and the digital output $t_{q1}$ can be expressed as:
\begin{eqnarray}
\label{eq4_noiseTF}
 \frac{t_{q1}}{t_{noise}} &=&  \frac{k_{t2i} Z_2(s) A_1(s) k_{v2t} }{1+A_1(s) k_{v2t} k_{t2i} Z_2(s)}\\
  &=&\frac{(1 + s r_o C_{fi})(1 + s r_{out} C_L)}{(1 + s r_o C_{fi})(1 + s r_{out} C_L) + 2 g_m r_o \frac{C_{fi}}{I_{cm}}r_{out} \frac{I_{on}}{T_0}} \nonumber
\end{eqnarray}
The transfer function in (\ref{eq4_noiseTF}) reveals second-order noise shaping, which effectively attenuates the quantization noise related to the DFF by the FI-DIGOTA DC gain, i.e. $A_0 \simeq 2 g_m r_o \frac{C_{fi}}{I_{cm}}r_{out} \frac{I_{on}}{T_0}$.

On the other hand, the NTF from the FI-DIGOTA equivalent input noise source $v_{n,in}$ to $t_{q1}$ can be expressed as:
\begin{eqnarray}
\label{eq4_noiseTF_input}
 \frac{t_{q1}}{v_{n,in}} &=&  \frac{A_1(s) k_{v2t} }{1+A_1(s) k_{v2t} k_{t2i} Z_2(s)}\\
&=& \frac{(1 + s r_{out} C_L) g_m r_o \frac{C_{fi}}{I_{cm}} }{(1 + s r_o C_{fi})(1 + s r_{out} C_L) + 2 g_m r_o \frac{C_{fi}}{I_{cm}}r_{out} \frac{I_{on}}{T_0}} \nonumber
\end{eqnarray}
Remarkably, (\ref{eq4_noiseTF_input}) reveals first-order noise shaping on the FI-DIGOTA input noise, and a low-frequency noise attenuation of $T_0/(2r_{out}I_{on})$.

\subsection{Noise Sources}


In the proposed model, the quantization noise $t$\textsubscript{noise} has been modeled as a discrete-time random process with a uniform amplitude distribution between $-T$\textsubscript{clk} and $T$\textsubscript{clk}, i.e. with a variance $T$\textsubscript{clk}\textsuperscript{2}/3 corresponding to the total noise power, and with a white power spectral density (PSD). It is important to note that, considering the FI-DIGOTA operation described above and illustrated in the waveforms of $v$\textsubscript{cmp}, $v$\textsubscript{outP} and $v$\textsubscript{outN} in Fig.\ref{Fig3_waveforms}, the  digital signals $OutP$, $OutN$ are inherently sampled at the FI-DIGOTA self-oscillation frequency $f_0=1/T_0$, which is significantly lower than the clock frequency. As a consequence, each noise sample is spaced in time by $T_0$ and thus, under the white noise assumption, the quantization noise power is evenly spread from DC to the Nyquist frequency $f_0/2$ and its PSD $S_{t_{noise}}$ can be estimated as:
\begin{equation}
\label{eq_nsd}
\int_{0}^{f_{0}} S_{t_{noise}}(f) \,df = \frac{T_{clk}^2}{3} \rightarrow S_{t_{noise}}(f) = \frac{T_{clk}^2 T_0}{3}
\end{equation}

On the other hand, in analogy with \cite{39_23_TOledo}, the FI-DIGOTA equivalent input noise is dominated by the shot noise contribution of the devices of the FI input stage, whose PSD can be expressed as:
\begin{equation}
\label{eq_vnsd}
S_{v_n,in}(f) = \frac{2qI_{cm}}{g_m^2}
\end{equation}
where $q$ is the elementary charge. Remarkably, the flicker noise contribution of the FI input stage is completely suppressed thanks to the dynamic opearation of the FI input stage \cite{Klumperink_rumoreFlickers}, resulting in competitive noise performance in the target biosensing application without requiring area- and power-hungry chopping or auto-zeroing techniques.

\subsection{Overall Noise and Noise Model Validation}

The overall noise spectrum referred to the digital output $t_{q1}$, and the individual contributions of the quantization noise and of the shot noise, calculated from the NFT derived in (\ref{eq4_noiseTF}) and (\ref{eq4_noiseTF_input}) and from the PSDs of the quantization noise and of the shot noise in (\ref{eq_nsd}) and (\ref{eq_vnsd}), respectively, is plotted in Fig.\ref{Fig6_NoiseTF} for the proposed DB-Potentiostat operated at 0.4 V $V$\textsubscript{dd}. 

The figure reveals that the quantization noise is dominant, and the behavior over frequency is characterized by +40\,dB/dec and +20\,dB/dec slopes, and a flat noise floor at low frequency.

In the same figure, the noise spectra calculated by the model are compared to the digital output noise power spectrum measured for the fabricated potentiostat at $V$\textsubscript{dd} = 0.4\,V with reference to the test setup described in Sect.V under zero input signal.

The curves are in reasonably good agreement at low frequency and up to the Nyquist frequency related to the inherent sampling at the FI-DIGOTA self-oscillation frequency (about 3kHz), in which the hypotheses of the average DIGOTA model in \cite{39_23_TOledo} are clearly no longer valid, and a peak is observed. On the other hand, the deviations at low frequency can be related to the inherent approximations of the same model  under dynamic conditions. Finally, the measured noise spectrum reveals no $1/f$ component at low frequency, confirming the inherent flicker noise suppression of the proposed DB-potentiostat.

\section{Mesoporous Microelectrode Design and Fabrication}

\begin{figure}[t]
  \centering
  \begin{subfigure}[b]{0.5\textwidth}
    \centering
    \includegraphics[width=\textwidth]{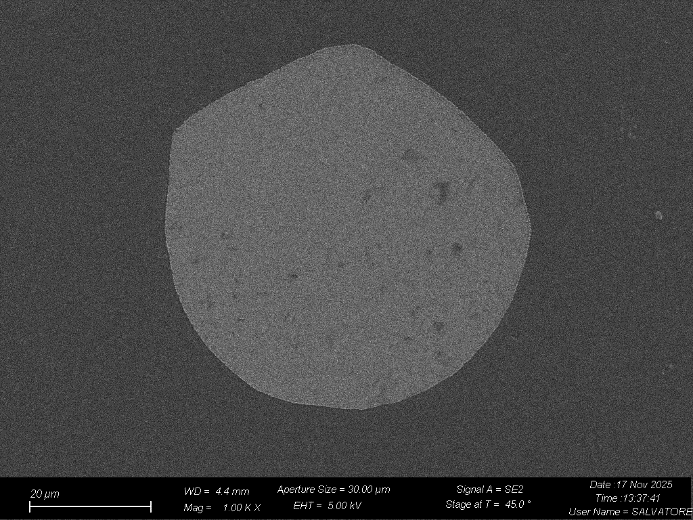}
    \caption{}
    \label{Fig7a_SEM}
  \end{subfigure}
  \begin{subfigure}[b]{0.5\textwidth}
    \centering
    \includegraphics[width=\textwidth]{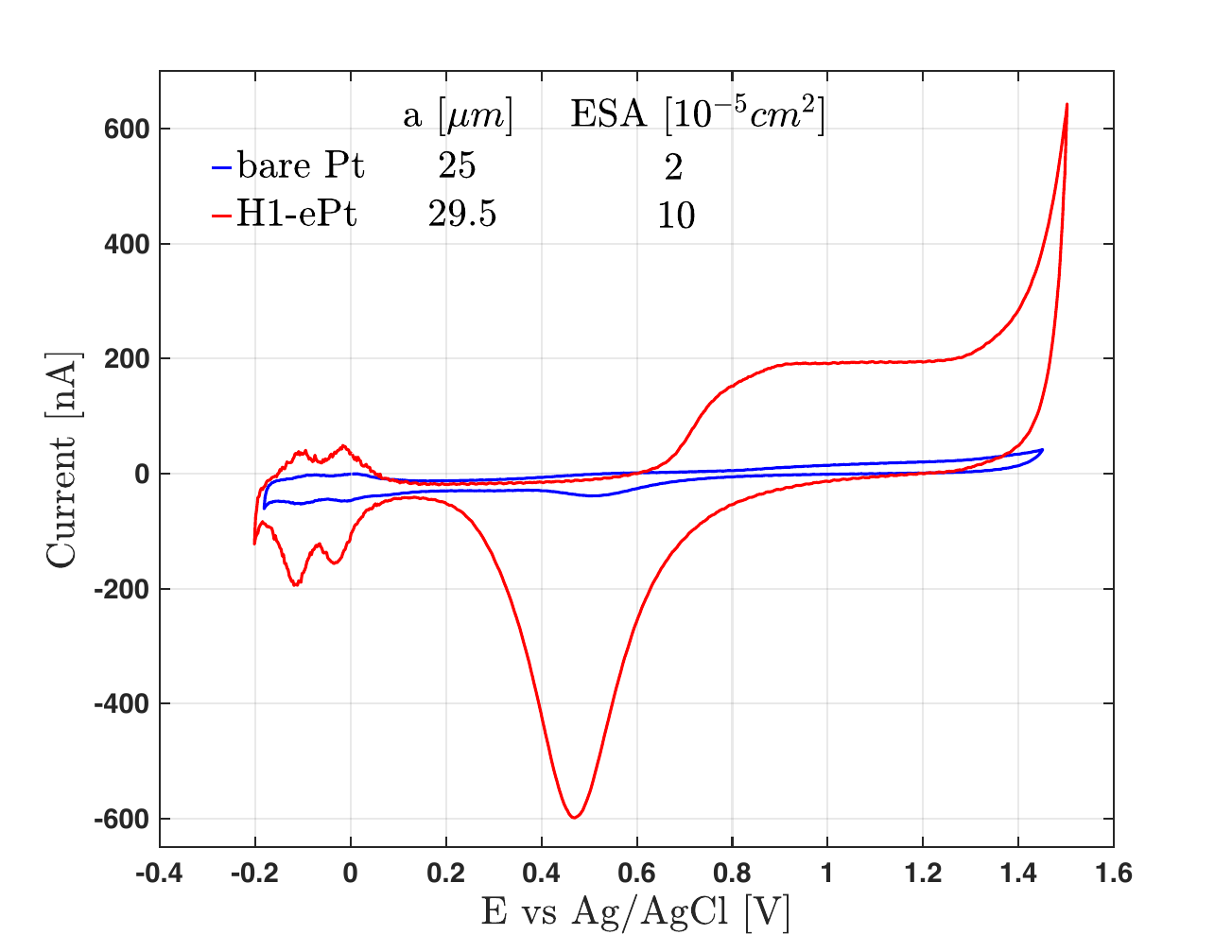}
    \caption{}
    \label{Fig7b_CvAcido}
  \end{subfigure}
  \caption{microelectrode characterization: (a) Scanning Electron Micrograph of the microdisc embedded in glass with H1-ePt performed by a ZEISS Supra 40 and Merlin Field Emission Scanning Electron Microscope (FESEM), (b) Cyclic voltammetry (CV) in acid with and without H1-ePt }
  \label{Fig7_ElectrodeCharacterization}
\end{figure}

As a crucial element to enable non-enzymatic glucose detection at nW-power, the electrodes of the electrochemical cell and their co-design with the electronic readout is detailed in this section. Initially, the choice of microelectrodes and nanostructuration process is justified, and then, the fabrication process is detailed. Finally, the electron transfer function of the microelectrodes is derived to relate the concentration of the analyte to the reaction current.



\subsection{Microelectrode Design and Fabrication}

Starting from \cite{11_34_Vassilyev1985PartI}\cite{12_35_Vassilyev1985_PartII}, several strategies have been explored to determine glucose concentration without the use of enzymes, a trend that is indeed still actual when coupled with the sensor miniaturization. 

These approaches primarily involve the design of various electrodes aimed at enhancing sensitivity to glucose oxidation while minimizing the selectivity of unmodified electrodes toward interfering analytes (i.e. L-ascorbic acid (AA) and uric acid (UA)). For this purpose, nanostructured glucose sensors have been reported in the literature with their  electrochemical characterization \cite{16_49_KHAMSAVI}\cite{17_50_Taurino2015}; however, to the best of our knowledge, they have never been designed targeting the constraints of real wearable applications. In this work, instead, the electrode is co-designed with the DB-potentiostat focusing on miniaturized dimensions (microelectrode), nA-range faradaic current under physiological glucose concentration, and operation at sub-500mV voltages.

Starting from these requirements, a Pt nanostructured electrode for non-enzymatic detection of glucose has been designed based on the hypothesis that the structural roughness within nanopores and mesopores is smaller than the characteristic scale of the diffusion field formed during a potential step.  Under these conditions,  reactants within the mesopores are consumed under diffusion-controlled regimes, resulting in faradaic currents that scale with the electrode’s geometric area for fast electron transfer reactions. Conversely, for reactions governed by kinetic control, the faradic current becomes highly dependent on the nanostructured features of the electrode, scaling with the rough electroactive surface area. The deeper and denser the nanopores and mesopores, the larger the roughness factor of the electrode and the larger the faradaic currents even for slow reaction rates.



Thanks to the peculiar geometry of the mesopores this nanostructurization allows to increase of order of magnitudes the electrochemical surface area while almost unmodified the geometrical area \cite{14_37_Evans2002}.





Leveraging the fact that the  reaction potential of the targeted molecules is coherent with the ULV supply domain of the circuit, our aim is to exploit the high sensitivity and the low-current range offered by the nonostructurated microelectrode to monitor the electrochemical current of two fundamental electrochemical reactions that proceed under diffusion controlled and kinetically controlled rates: the oxidation of ferrocyanide to ferricyanide and the oxidation of glucose, respectively. Experimental results referring to both reactions performed with the DB-potentiostat will be presented in Sect.V. 

\subsection{Electrode Transfer Function for Microelectrodes}

Microelectrodes are generally used to measure rates of electron transfer of fast reactions \cite{47_45_pletcher2001} and microdisc-shaped electrodes exhibit stable spherical diffusion under quiescent conditions \cite{48_46_Fleischmann1987}. When embedded within an insulating material of sufficiently large radial dimensions, microdiscs are capable of achieving steady-state (limiting) currents under mass transport controlled reactions, which can be described by the following expression:
\begin{equation}
    i_L = 4 n F D c a
    \label{eq5_Imiceoelettrodi}
\end{equation}
where $n$, $F$, $D$ and $a$ denote the number of electrons, the Faraday constant, the diffusion coefficient and the electrode radius. Once the electrode geometry and reaction mechanism are defined, the bulk concentration of the analyte can be readily determined from the limiting current ($i$\textsubscript{L}) within sub-second to a few seconds, depending on factors such as temperature, solution viscosity, and electrode radius.

\subsection{Fabrication and Characterization of the Mesoporous Pt Film}

We fabricated mesoporous Pt microdiscs electrodes according to the method firstly reported by Attard et al. \cite{13_36_Attard} and used as hydrogen peroxide sensor by Evans \cite{14_37_Evans2002} and by the same authors in previous studies for the detection of dissolved oxygen and pH \cite{15_38_Serrapede2013}.

The microelectrodes have been fabricated starting from platinum microdisc electrodes by Metrohm with nominal radii of 5\,\textmu m, 25\,\textmu m, 50\,\textmu m and 100\,\textmu m. The electrodes were polished with multigrade SiC papers, diamond powder (1\,\textmu m) and alumina powder (0.03\,\textmu m) until mirror finishing. The real geometrical area was measured by means of FESEM micrographs. Electrochemical cleaning was carried out in 0.5\,M H\textsubscript{2}SO\textsubscript{4} between -0.45\,V and +1.00\,V vs Ag/AgCl until reproducible Cyclic Voltammetries (CV) were obtained. This cleaning was carried out before any experiment.

The mesoporous Pt film was electrodeposited from a hexagonal (H1) liquid crystalline phase composed of the non-ionic surfactant Brij 58 (0.42\,g), hydrogen hexachloroplatinate hydrate (0.29\,g), distilled water (0.29\,g) and two drops of heptane \cite{Mara2804} at a potential of -0.06\,V vs Ag/AgCl until the charge density of 4 \textmu C/cm\textsuperscript{2} was passed. The geometrical area was evaluated by means of FESEM micrographs and by the limiting current according to (\ref{eq5_Imiceoelettrodi}). The roughness factor was evaluated by using the cycling voltammetries of the electrochemical cleaning after the electrodeposition by measuring the areas under the hydrogen adsorption/desorption peaks of the cyclic voltammograms and using the conversion factor of 210\,\textmu C cm\textsuperscript{-2} to determine the active surface according to Trasatti’s method \cite{45_43_TRASATTI1992353} and corrected to Biegler’s observation \cite{46_44_BIEGLER1971269}.


Based on the developed model, the DB potentiostat is expected to track the faradaic current determined by the electrode impedance and reaction kinetics. The following measurements validate this behavior using both diffusion-controlled and kinetically-controlled electrochemical processes.

\section{Experimental Results}

Here, the measurement setup is shown for both electrical and chemical components. Then, the DB potentiostat readout capabilities are tested in diffusion control regime, for the ferrocyanide sensing.

\begin{figure}[t]  
    \centering
    \includegraphics[width=0.5\textwidth]{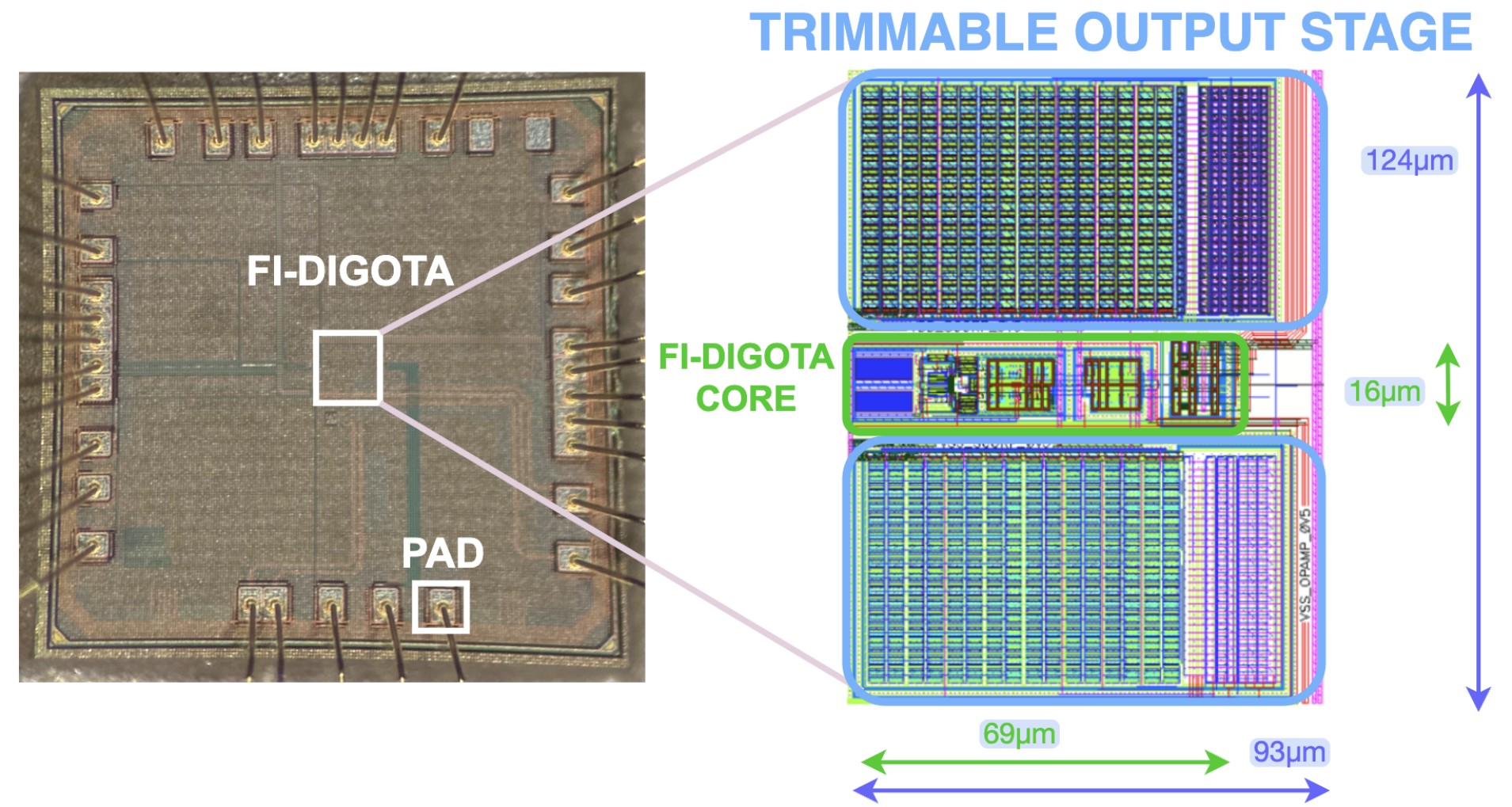} 
        \caption{FI-DIGOTA micrograph and layout.}
   \label{Fig8_Layout}
\end{figure}

\begin{figure}[t]
  \centering
  \begin{subfigure}[b]{0.5\textwidth}
    \centering
    \includegraphics[width=\textwidth]{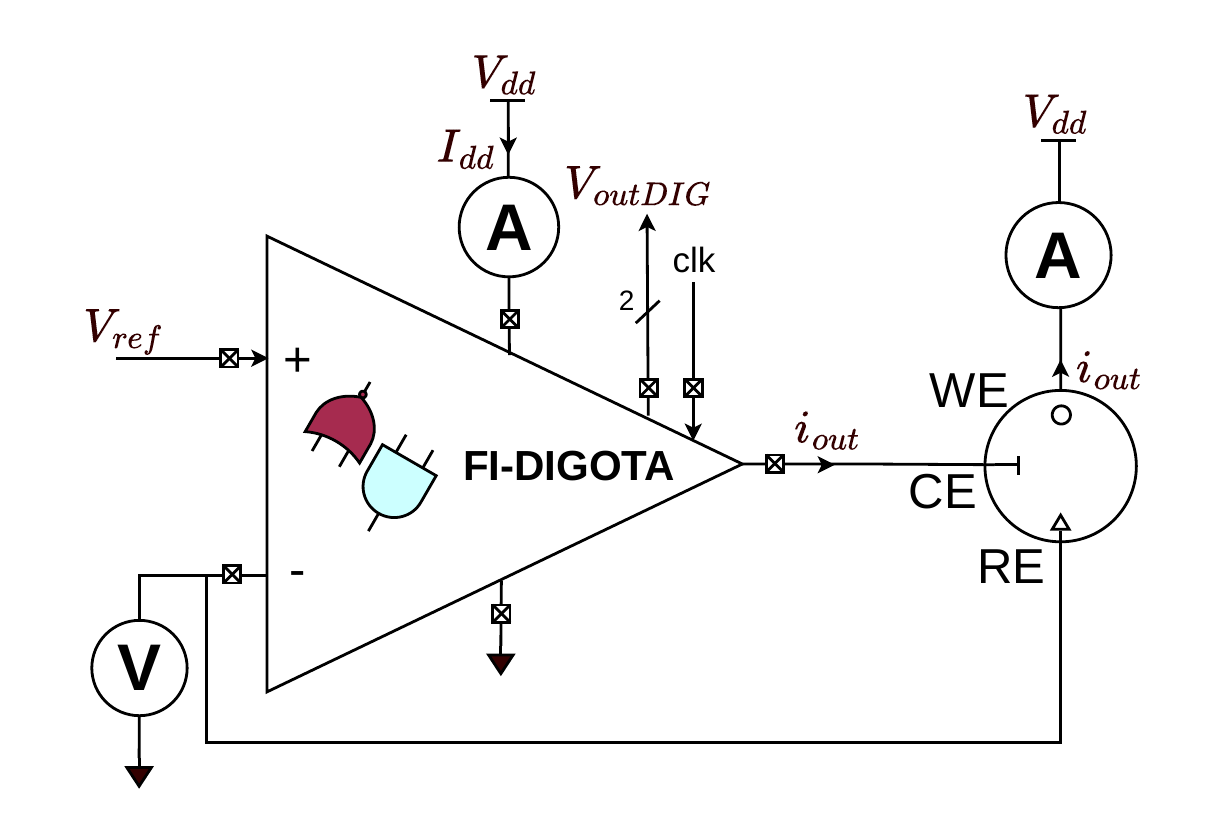}
    \caption{}
  \end{subfigure}
  \begin{subfigure}[b]{0.5\textwidth}
    \centering
    \includegraphics[width=\textwidth]{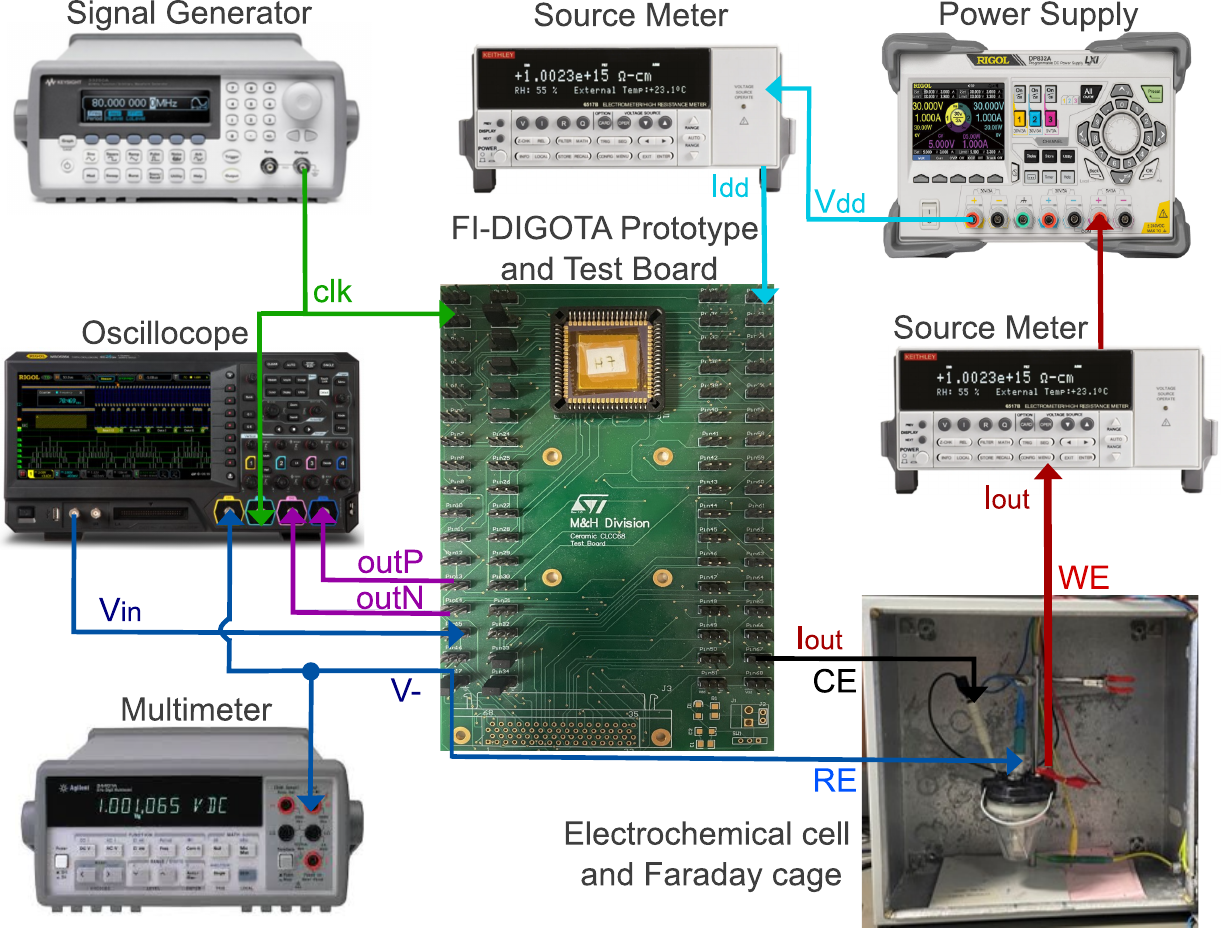}
    \caption{}
  \end{subfigure}
  \caption{Measurement setup for electrochemical current sensing: (a) schematic, (b) block diagram.}
  \label{Fig9_MeasurementSetup}
\end{figure}

\subsection{Experimental test setup}
Taking advantage of the signal and noise analysis presented in Sect.III and from an estimate of the expected faradaic current based on (\ref{eq5_Imiceoelettrodi}), the FI-DIGOTA has been designed and fabricated in 130\,nm CMOS technology. The layout and the micrograph of the circuit are shown in Fig.\ref{Fig8_Layout}. The FI-DIGOTA occupies a total area of 11,532 \textmu m\textsuperscript{2}, with more than 90\% of the area allocated to the trimmable output stage.

The measurement setup is depicted in Fig.\ref{Fig9_MeasurementSetup}. 
For the electrical measurement the following instruments have been used: source meter Keithley 6517B, multimeter Agilent 34401A, signal generator Agilent 33250A, DC power supply Rigol DP832A and, Oscilloscope Rigol MSO5104.
Two source meters Keithley 6517B have been used to monitor the electrochemical and the supply currents while, the digital pulses ($v$\textsubscript{outP},$v$\textsubscript{outP}) were recorded using the oscilloscope and then processed in Matlab to evaluate the faradic current through equation (\ref{eq1_FaradicCurrent}).

Experimental results supporting the frequency-domain signal and noise analysis of the DB-potentiostat presented in Sect.III have been also obtained from the above test setup, while a complete characterization of the FI-DIGOTA as a standalone amplifier is detailed in \cite{40_24_DeGregorio}.

Electrochemical experiments were all performed inside a grounded Faraday cage. The control experiments and the characterization of the electrodes were carried out with a potentiostat/galvanostat Metrohm PGSTAT304N and Nova 2.0. Depending on the cell design, the reference electrode was Ag/AgCl\textsubscript{sat NaCl} or a pseudo-reference (Pt coil) immersed in the testing solution. As counter electrode a platinum bar was used and the microdisc electrodes described in Sect.IV were employed as working electrodes.

\subsection{FI-DIGOTA Current Readout Multidie Characterization}

Before proceeding with the validation of the potentiostat using electrochemical reactions and, in order to verify the reliability of the FI-DIGOTA as a current readout circuit under process variations, five dice were tested with a resistive load in series to a regulated DC power supply at the output instead of the electrochemical cell, so that to set the FI-DIGOTA output current by varying the DC voltage of the power supply. The measurements were carried out by externally varying the FI-DIGOTA output current across more than two orders of magnitude, and estimating the FI-DIGOTA current from the digital output by (\ref{eq1_FaradicCurrent}).

\begin{figure}[t]  
    \centering
    \includegraphics[width=0.5\textwidth]{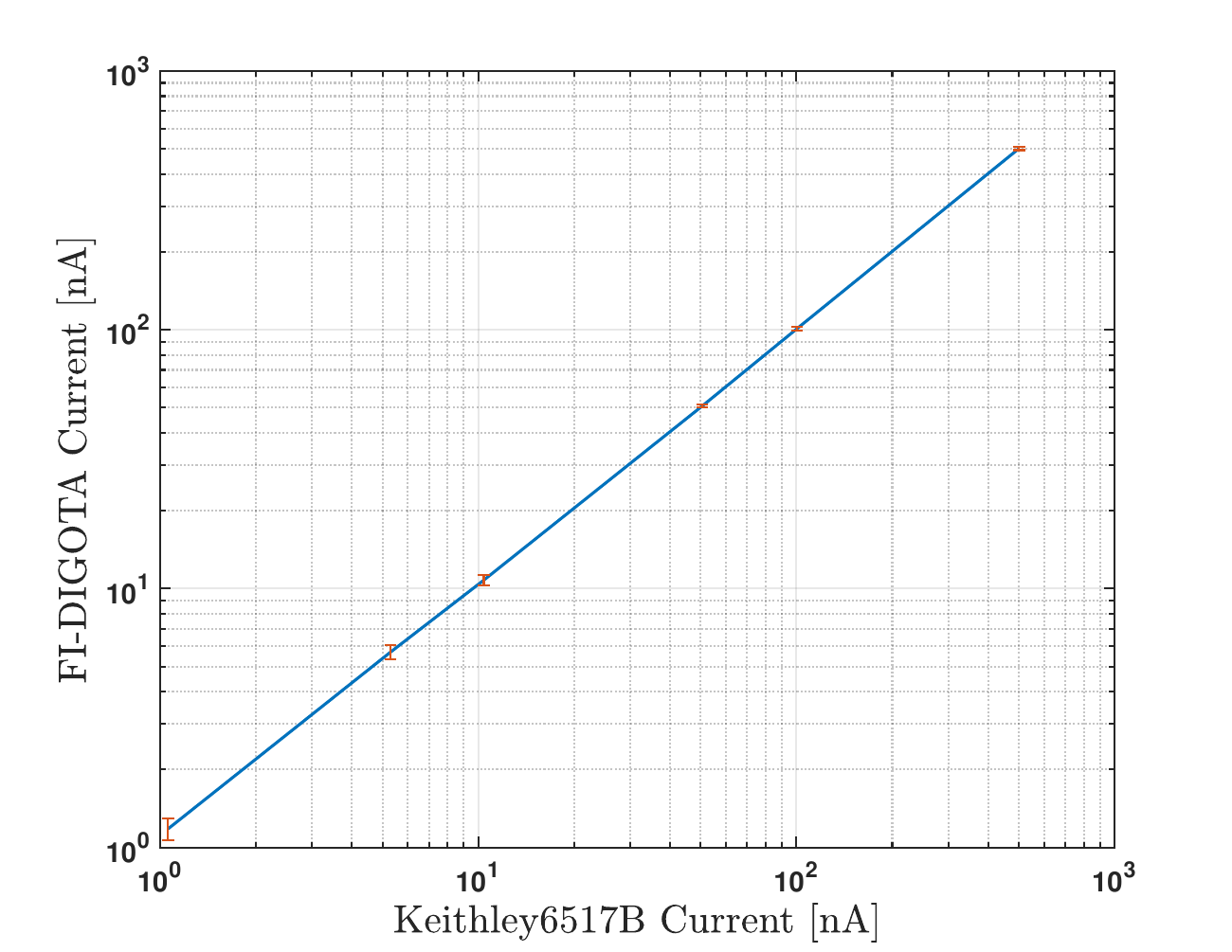} 
        \caption{Multi-die FI-DIGOTA characterization as a current readout with resistive load.}
   \label{Fig12_MultiChipCorrente}
\end{figure}

The obtained results are shown in Fig.\ref{Fig12_MultiChipCorrente} listed in table \ref{table3}, demonstrating consistent current readout across five dice. In detail, the average standard deviation across five dice of the readout current normalized to the current read with the picoamperometer, turns out to be equal to 4.46\,\% and in line with the requirements of PoC diagnostics.

\begin{table}
\centering
\caption{FI-DIGOTA Multidie Current Measurement}
\label{table3}
\setlength{\tabcolsep}{3pt}
\begin{tabular}{|c|c|c|c|c|}
\hline
 $V$\textsubscript{dd} [V]& Mean [nA] & $\sigma$ [nA] & Min [nA] & Max [nA] \\ \hline
  \multirow{2}{*}{0.3}& 1.18 & 0.12 & 1.08 & 1.37 \\ \cline{2-5}
  & 5.68 & 0.35 & 5.25 & 6.21 \\ \hline
    \multirow{2}{*}{0.4}& 10.79 & 0.48 & 10.12 & 11.36 \\ \cline{2-5}
  & 50.77 & 0.76 & 49.67 & 51.58 \\ \hline
    \multirow{2}{*}{0.5}& 100.99 & 1.64 & 98.23 & 102.41 \\ \cline{2-5}
  & 502.28 & 7.75 & 490.14 & 511.47 \\ \hline

\end{tabular}
\end{table}

\subsection{Amperometric detection in a diffusion-controlled regime: the ferrocyianide sensing}

In this section, the amperometric detection of the oxidation reaction of ferrocyanide is analyzed. Initially, the nanostructured micro-electrode has been characterized using commercial instruments, while subsequently, the DB potentiostat has been employed to measure the reaction current.

\subsubsection{Electrode Characterization}
The chronoamperograms were first acquired with PGSTAT304N to estimate the time required for the steady state current to occur and then the potential step was done with the DB Potentiostat. With DB Potentiostat the reaction current was acquired for 5\,s, then averaged according to (\ref{eq1_FaradicCurrent}) and compared with that measured with PGSTAT304N in the same interval.

\begin{figure}[htbp]
  \centering
    \begin{subfigure}[b]{0.48\textwidth}
    \centering
    \includegraphics[width=\textwidth]{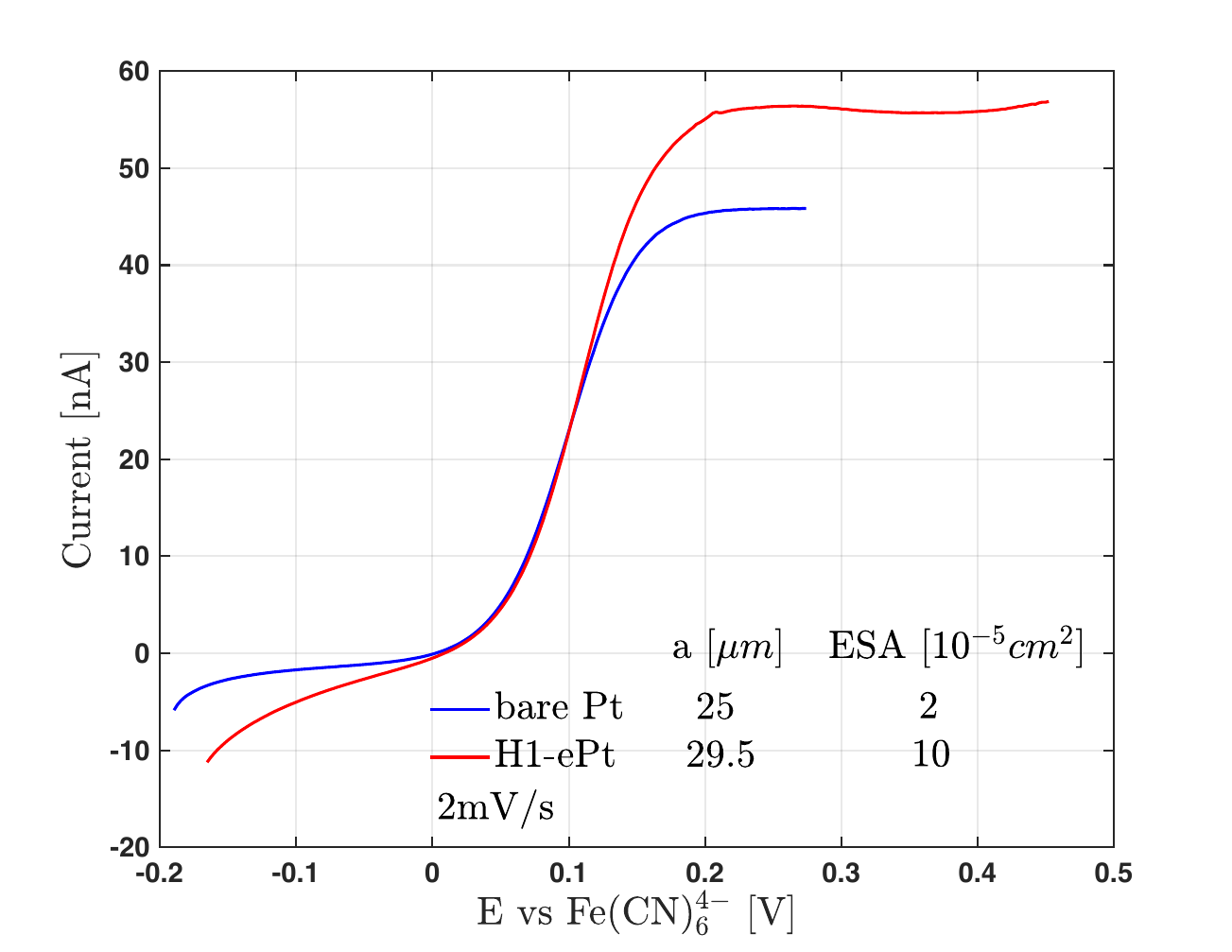}
    \label{Fig10_CVelectrode}
      \end{subfigure}
     \vspace{-0.3cm}

  \caption{Fe(CN)\textsubscript{6}\textsuperscript{4-} Linear Sweep Voltammetry (LSV) at 2\,mV/s scan rate with and without H1-ePt}
  \label{Fig10_ElectrodeCharacterization}
  \vspace{-0.3cm}
  \begin{subfigure}[b]{0.48\textwidth}
    \centering
    \includegraphics[width=\textwidth]{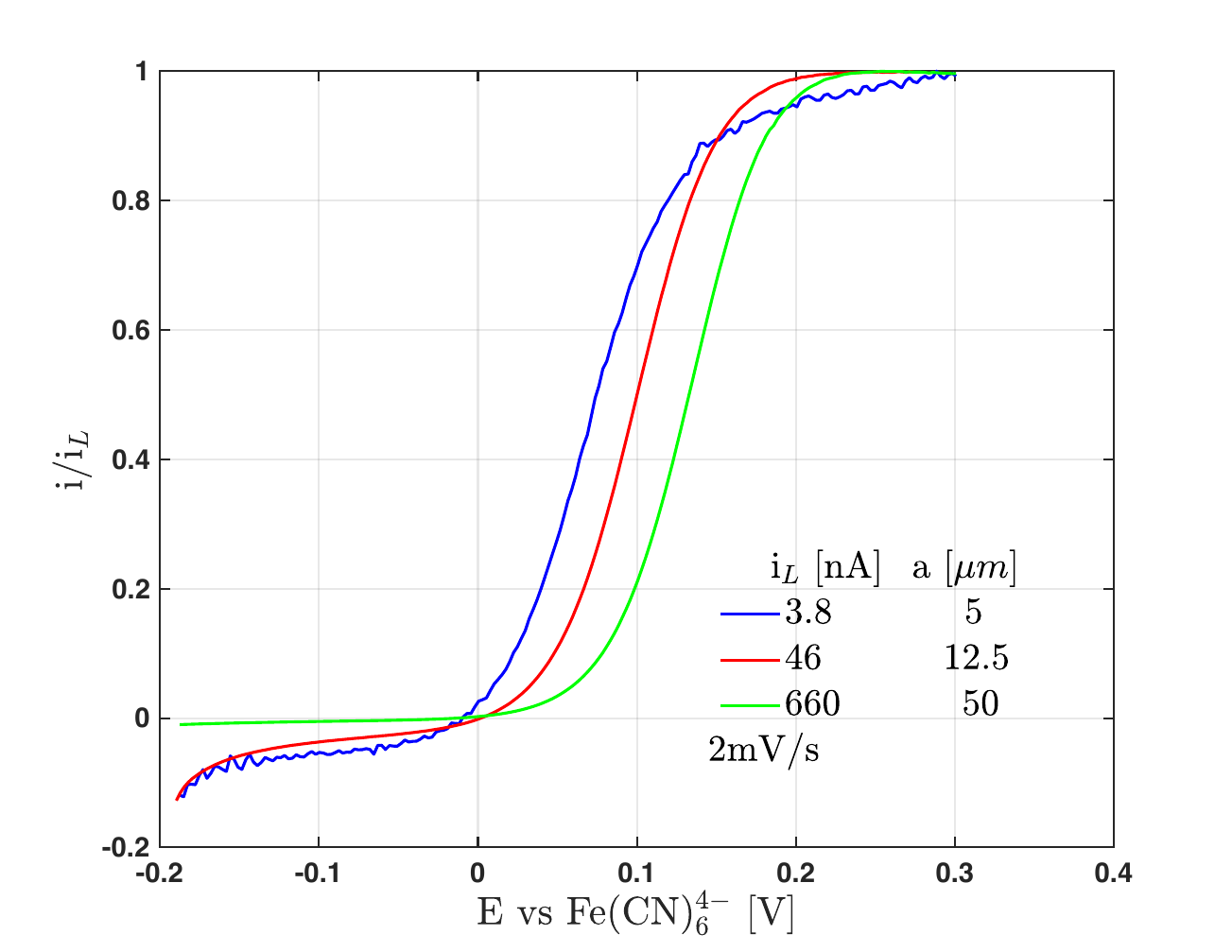}
    \caption{}
      \vspace{-0.3cm}

      \label{Fig11a_LSV_ferrocianato}
  \end{subfigure}
  \begin{subfigure}[b]{0.48\textwidth}
    \centering
    \includegraphics[width=\textwidth]{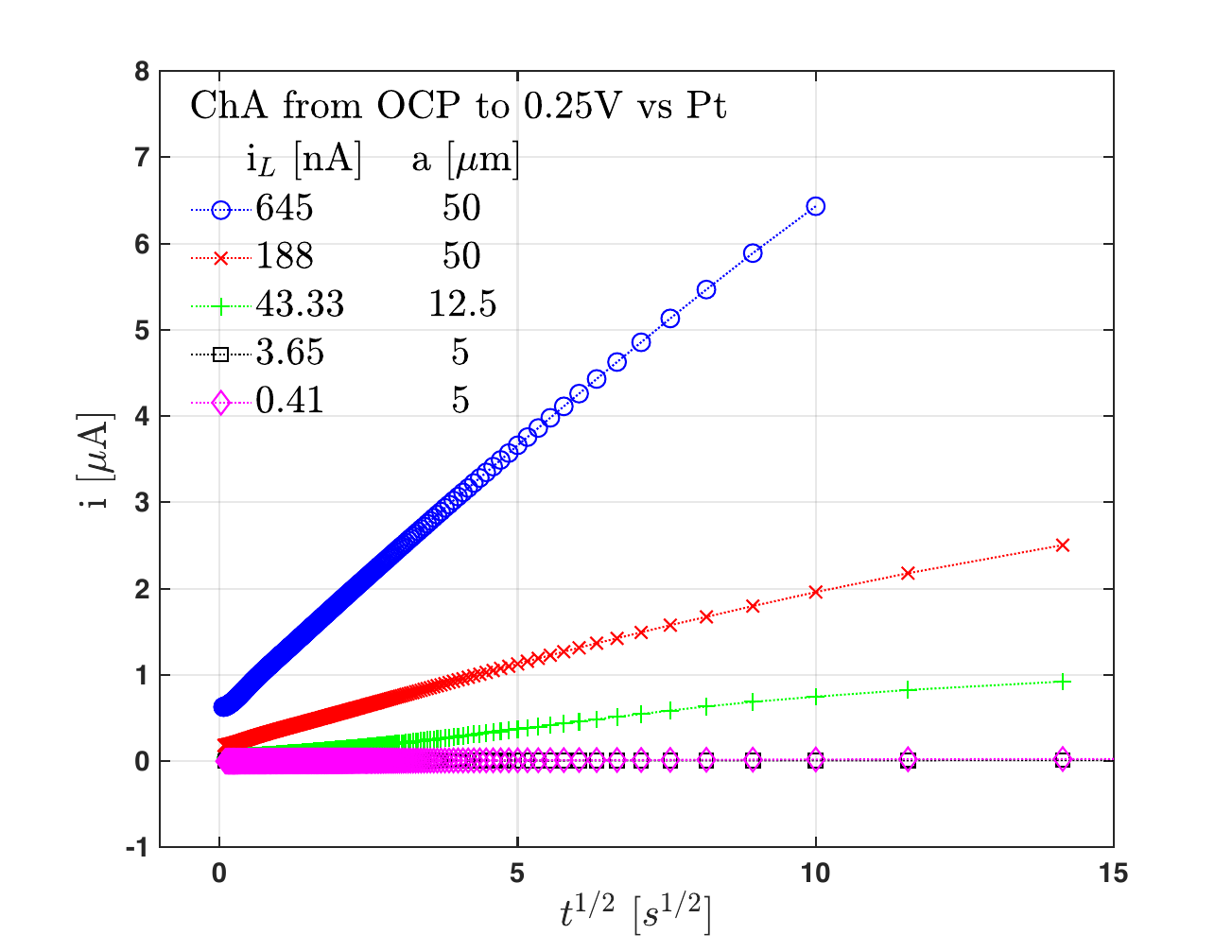}
    \caption{}
          \label{Fig11b_chronoFerrocianato}
  \end{subfigure}
  \caption{Electrode characterization for ferrocyanide sensing: (a) Linear Sweep Voltammetry (LSV) at 2 mV/s scan rate acquired on micro-discs with different radius, (b) Chronoamperograms from open circuit potential (OCP) to 0.25\,V vs Pt and fit according to Mahon-Oldham.}
  \label{Fig8}
\end{figure}

\begin{figure}[t]  
    \centering
    \includegraphics[width=0.5\textwidth]{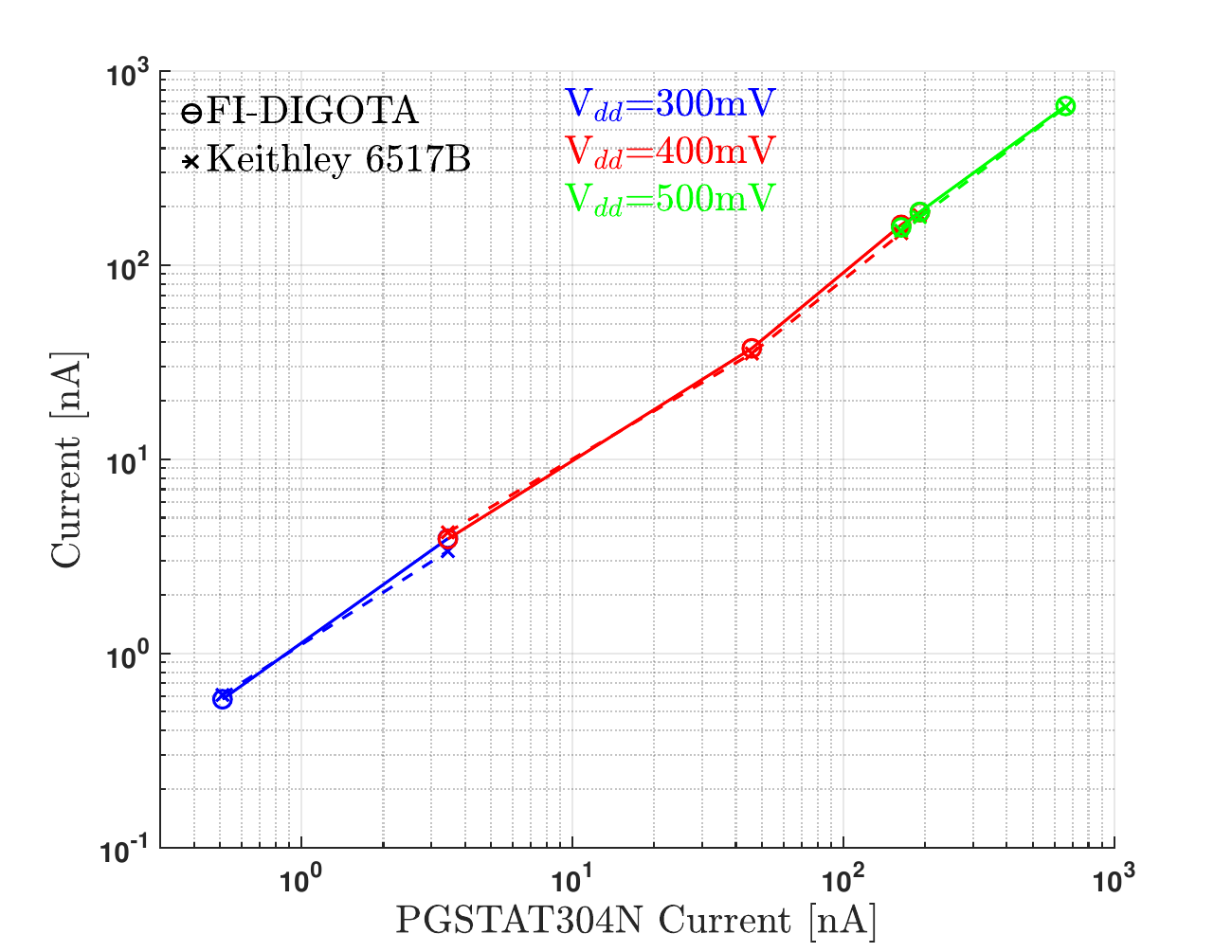} 
        \caption{FI-DIGOTA ferrocyanide characteristic.}
   \label{Fig13_DIGOTAvsPGSTAT_ferrocianto}
\vspace{-20pt}
\end{figure}

\begin{figure}[htbp]
  \centering
  \begin{subfigure}[b]{0.45\textwidth}
    \centering
    \includegraphics[width=\textwidth]{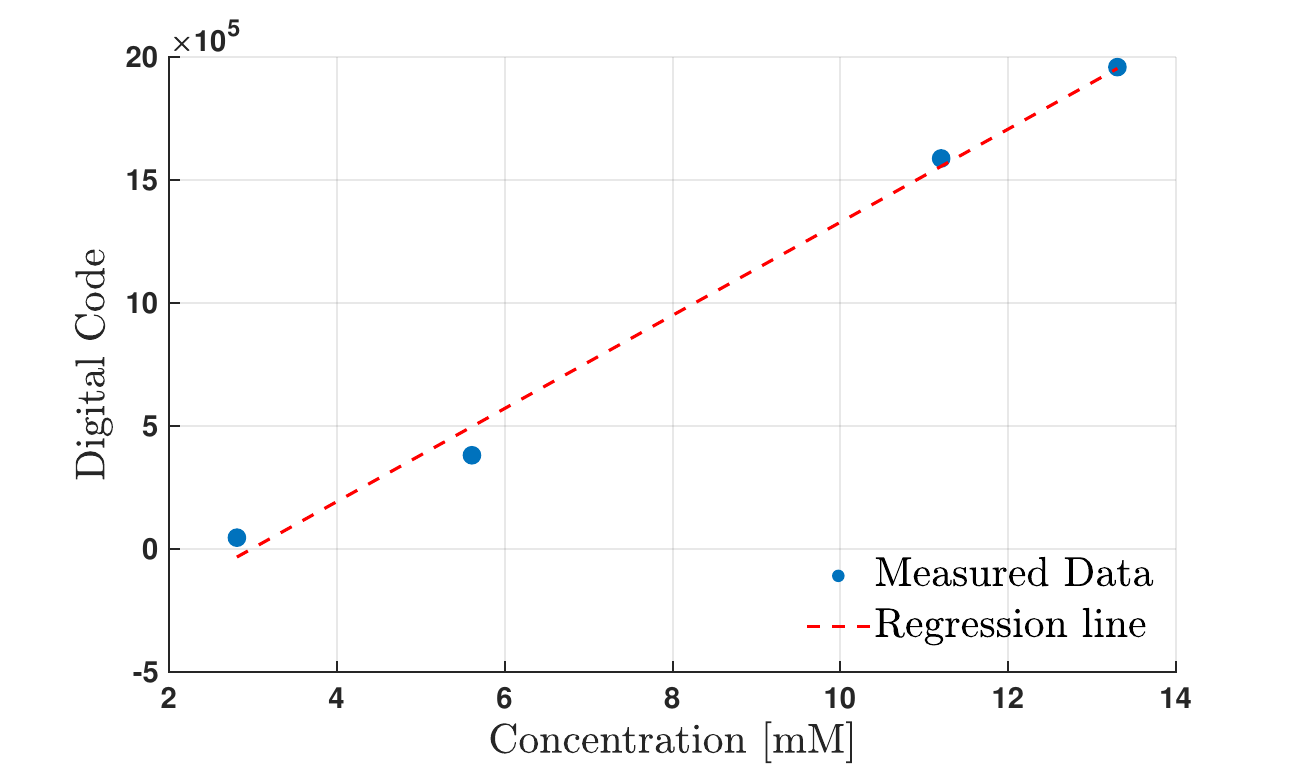}
    \caption{}
  \end{subfigure}
  \begin{subfigure}[b]{0.45\textwidth}
    \centering
    \includegraphics[width=\textwidth]{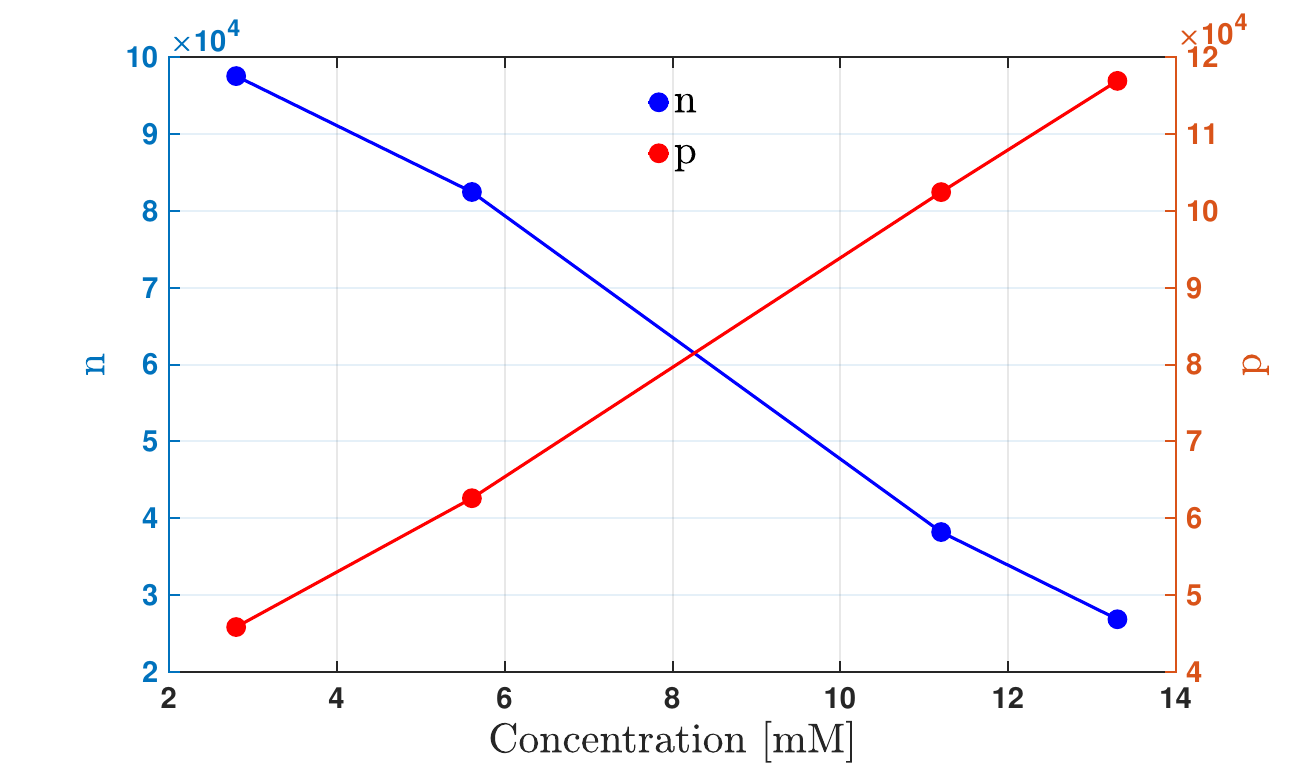}
    \caption{}
  \end{subfigure}
  \begin{subfigure}[b]{0.45\textwidth}
    \centering
    \includegraphics[width=\textwidth]{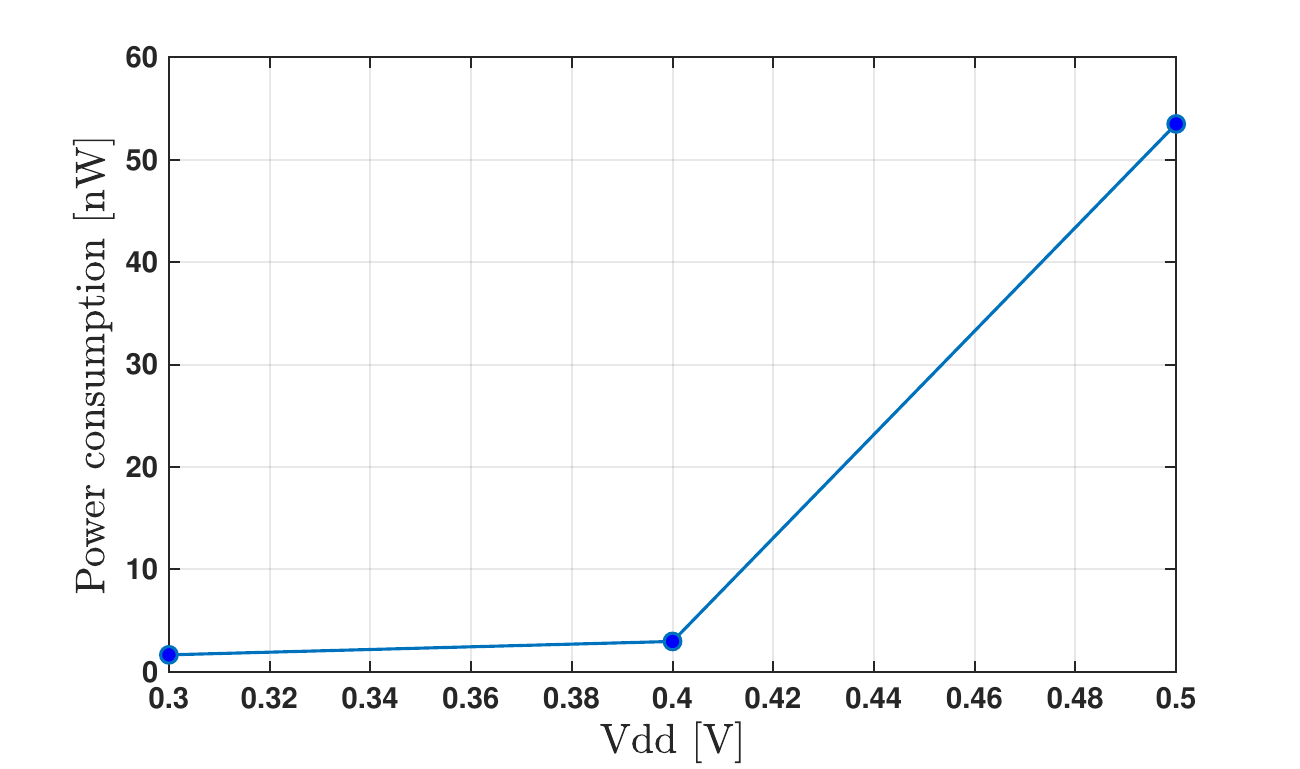}
    \caption{}
  \end{subfigure}
    \begin{subfigure}[b]{0.45\textwidth}
    \centering
    \includegraphics[width=\textwidth]{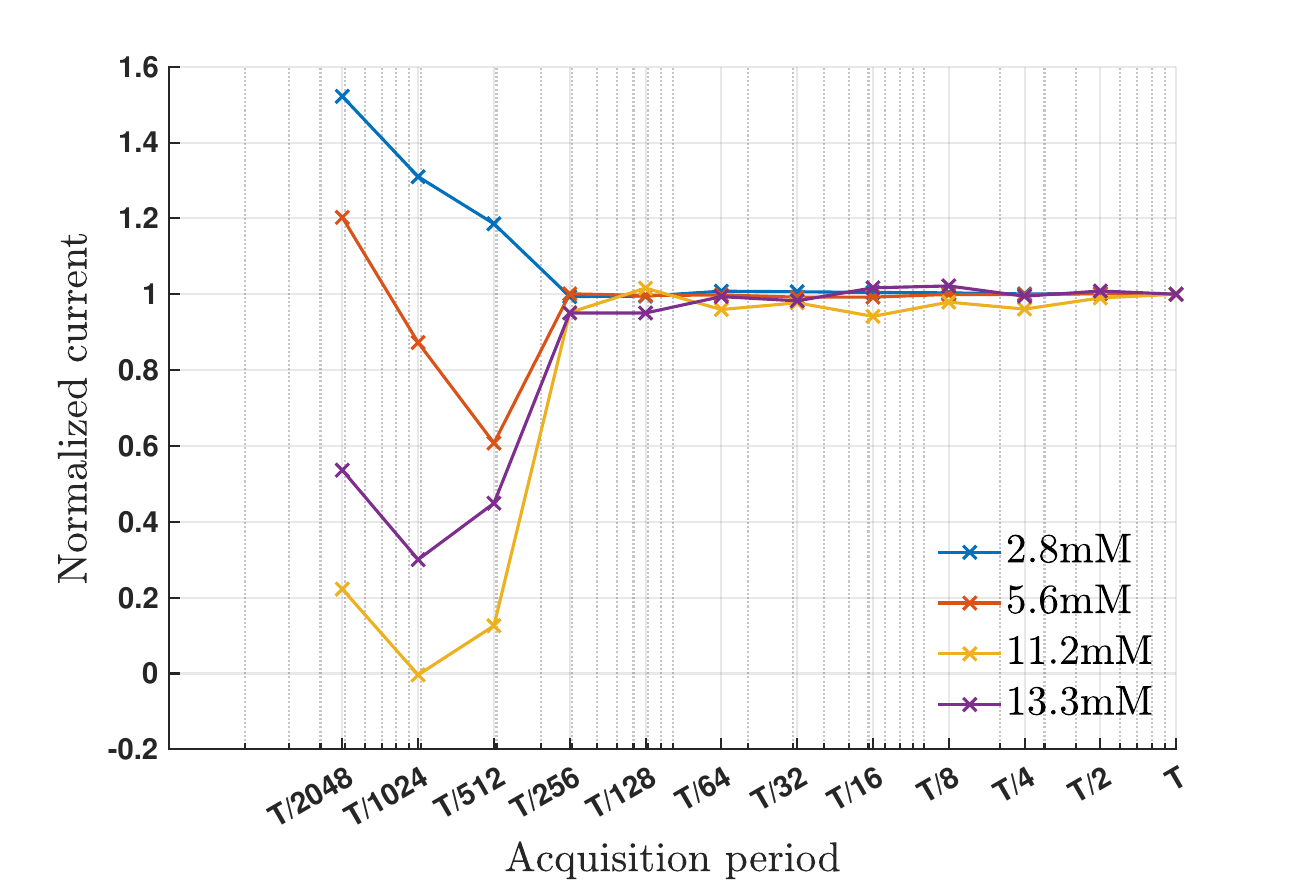}
    \caption{}
  \end{subfigure}
  \caption{DB potentiostat electrical behavior with $V$\textsubscript{dd} = 400\,mV for ferrycindide oxidation measurements during 5\,s acquisitions: (a) Sensitivity, (b) Digital pulses evolution, (c) Power consumption, (d) Minimum acquisition time.}
  \label{Fig14}
\end{figure}

As shown in Fig.\ref{Fig10_ElectrodeCharacterization} the Linear Sweep Voltammetry (LSV) of a microdisc electrode recorded in ferrocyanide solution using both mesoporous and  bare Pt electrodes displayed very similar electrochemical behavior. In fact, they showed comparable electrochemical reversibility (both with a slope of -24\,mV) and limiting current normalized by geometric area.

These similarities in diffusion-controlled behavior between the mesoporous and bare Pt electrodes are consistent with previous findings \cite{Estrele_added}.

The steady-state current response of Pt microdisc electrodes in Na\textsubscript{4}Fe(CN)\textsubscript{6} was used to evaluate the operating range of the DB potentiostat across four orders of magnitude (from hundreds of pA to hundreds of nA). To span the desired current range, three electrodes with different radii and several solution concentrations of Na\textsubscript{4}Fe(CN)\textsubscript{6} 0.5\,M NaCl have been empolyed. The characterization of three sensors is reported in Fig.\ref{Fig11a_LSV_ferrocianato} were it is possible to observe that when normalizing the current with their respective $i$\textsubscript{L}, a shift in potential due to kinetic limitations become evident: the smaller the electrode radius, the more negative the overpotential and the half-wave potential.

Chronoamperometry experiments were performed in several solutions of different concentration of Na\textsubscript{4}Fe(CN)\textsubscript{6} by a potential step from OCP to +250\,mV vs OCP for 120\,s in accordance with the voltammetries in Fig.\ref{Fig11a_LSV_ferrocianato}. Fig.\ref{Fig11b_chronoFerrocianato} shows such current transients obtained and compared with the fittings to the Mahon-Oldham equation (dashed lines)\cite{49_47_Mahon2005}. For all the electrode radii, the fittings agree at long times, indicating that the electrodes follow the expected behavior of a microdisc. At short times, data are not perfectly in line with the theoretical curves probably due to some adsorbed oxygen on the Pt electrode because the solution was not degassed. 
The current transient curves after the potential step from OCP to +250\,mV were acquired either by PGSTAT302N, either by the FI-DIGOTA and the Keithley6517B.

\subsubsection{DB Potentiostat Measurements}
For all the measurements, the DB Potentiostat clock frequency has been kept equal to 50\,kHz, while the strength of the output stage has been chosen accordingly to the expected electrochemical current range (\ref{eq5_Imiceoelettrodi}).
Since the ferrocyianide oxidation foresees a voltage at the CE higher than the supply voltage value ($v$\textsubscript{ce} $>$ $V$\textsubscript{dd}), an external regulated DC power supply is used.
The current measured after 100\,s from the potential step was acquired by the three systems for 5\,s and the comparative data are reported in Fig.\ref{Fig13_DIGOTAvsPGSTAT_ferrocianto}.  

From this dataset it is clear that DB potentiostat and commercial instruments (Keithley6517B and PGSTAT302N) have identical current responses over more than three decades of current from hundreds of pA to hundreds of nA. To detect the reaction current in several orders of magnitude, different combinations of $V$\textsubscript{dd} and static current ($i$\textsubscript{p}, $i$\textsubscript{n}) have been used as shown in Fig.\ref{Fig13_DIGOTAvsPGSTAT_ferrocianto}. 

The circuit exhibits an high linearity of $R^2$ = 0.991 and, in Fig.\ref{Fig14}a the regression line for the measurement with $V$\textsubscript{dd}= 400mV is shown. Since the minimum digitally representable current (LSB) is defined by a single current pulse of duration $T$\textsubscript{clk} and amplitude $i$\textsubscript{p} (pull-up current) divided by the number of clock cycles of the acquisition (M), it is possible to express the faradic currents with an associated digital code. 

The DB potentiostat sensitivity will thus depend on the acquisition period, clock frequency and, on the supply voltage, reaching a maximum sensitivity of 334,137\,LSB/mM, exploiting a $V$\textsubscript{dd} = 300\,mV, FI-DIGOTA static currents $i$\textsubscript{p} = 4.89\,nA and $i$\textsubscript{n} = 10.16\,nA, and an acquisition period equal to 5\,s.
Fig.\ref{Fig14}b shows the trend of the digital pulses activations (n and p, number of times signals $v$\textsubscript{outP} and $v$\textsubscript{outN} are active in the acquisition interval), and reveals that these parameters follow (\ref{eq1_FaradicCurrent}) as function of the analyte concentration. 

The power consumption, as expected in subthreshold regime, exponentially increases with the supply voltage, as shown by Fig.\ref{Fig14}c. The minimum power consumption is obtained with a supply voltage of 0.3\,V and is equal to 1.65\,nW.

The input referred noise has been evaluated with different $V$\textsubscript{dd}, averaging the digital output streams for the entire acquisition period. The obtained values are equal to 5.6\,pA\textsubscript{rms}, 46.72 \,pA\textsubscript{rms} and 347.03 \,pA\textsubscript{rms} for $V$\textsubscript{dd} = 0.3\,V, 0.4\,V, 0.5\,V, respectively. Based on that, the dynamic ranges are evaluated as function of $V$\textsubscript{dd} and considering the current range measured in Fig.\ref{Fig13_DIGOTAvsPGSTAT_ferrocianto}, giving back values of 53.82\,dB, 71.46\,dB, 63.20\,dB for $V$\textsubscript{dd} equalt to 300mV, 400mV and 500mV, respectively. The variability in dynamic range is solely due to the different current ranges employed. Notably, since the input noise current increases exponentially in the subthreshold regime, and considering a current range spanning one order of magnitude for each V\textsubscript{dd}, a constant dynamic range of 59 dB is achieved. Furthermore, for all V\textsubscript{dd} values, the trimming circuit can be utilized to enhance the circuit's dynamic range.

Finally, Fig.\ref{Fig14}d defines the minimum acquisition time, showing the reaction current normalized with respect to its steady-state value, whose estimate is obtained by considering a limited set of samples acquired by the DB potentiostat over $T=5$s, as well as by considering the full set of acquired data.
From the figure, an acquisition time of $T/256$=20 ms is sufficient to obtain a good estimate of the faradaic current. This is consistent with the noise analysis presented in Sect.III and with the results in Fig.\ref{Fig6_NoiseTF}.



\section{Non-Enzymatic Glucose Detection Proof of Concept}

In this section the measurement results for the DB potentiostat experimental characterization in the target non-enzymatic glucose monitoring application, which involves the glucose oxidation under a kinetically-controlled regime, are shown.


\subsection{Amperometric detection in a kinetically-controlled regime: the glucose sensing}

Here, the characterization of the nanostructured micro-electrode for non-enzymatic glucose detection is shown and, subsequently, the measurement results obtained with the DB potentiostat are presented.

\subsubsection{Electrode Characterization}

\begin{figure}[t]
  \centering
  \begin{subfigure}[b]{0.5\textwidth}
    \centering
    \includegraphics[width=\textwidth]{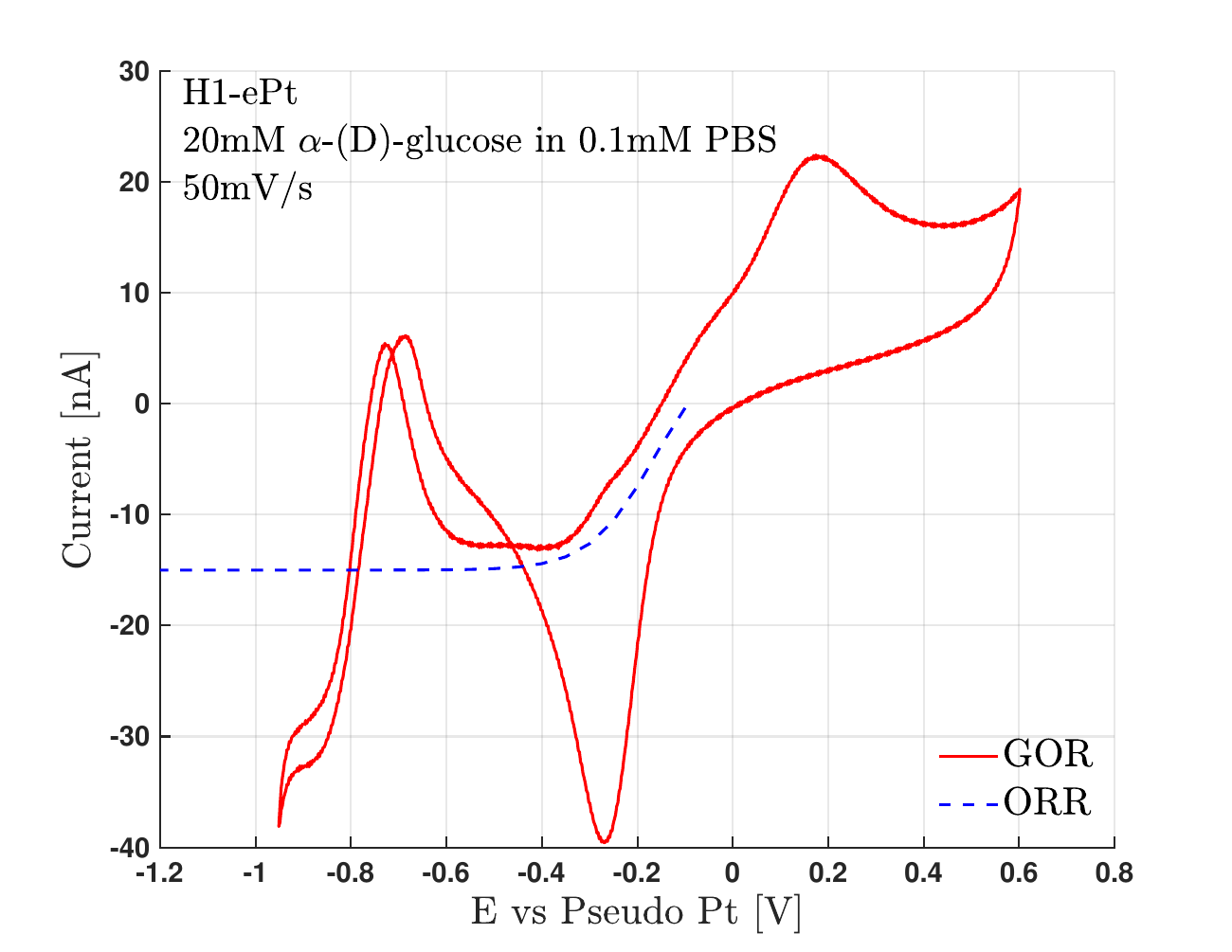}
    \caption{}
      \label{Fig15a_CVnanostructure}
  \end{subfigure}
  \begin{subfigure}[b]{0.5\textwidth}
    \centering
    \includegraphics[width=\textwidth]{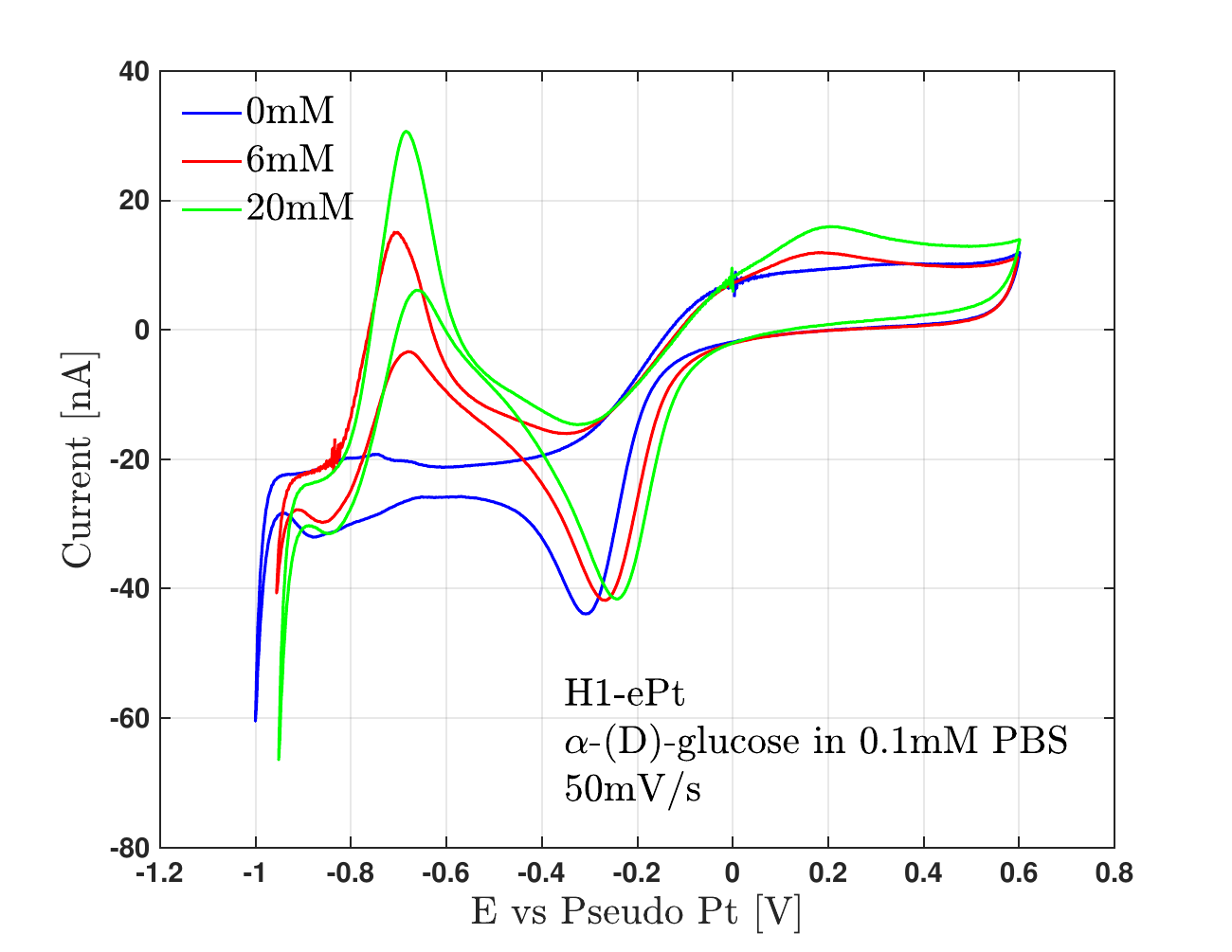}
    \caption{}
      \label{Fig15b_LSVglucose}
  \end{subfigure}
  \caption{Setup for electrochemical current sensing: (a) Cyclic Voltammetry (CV) with H1ePt nanostructure, (b) Cyclic Voltammetry (CV) in 0.1M PBS with different glucose concentrations.}
  \label{Fig11_}
\end{figure}

\begin{figure}[t]  
    \centering
    \includegraphics[width=0.5\textwidth]{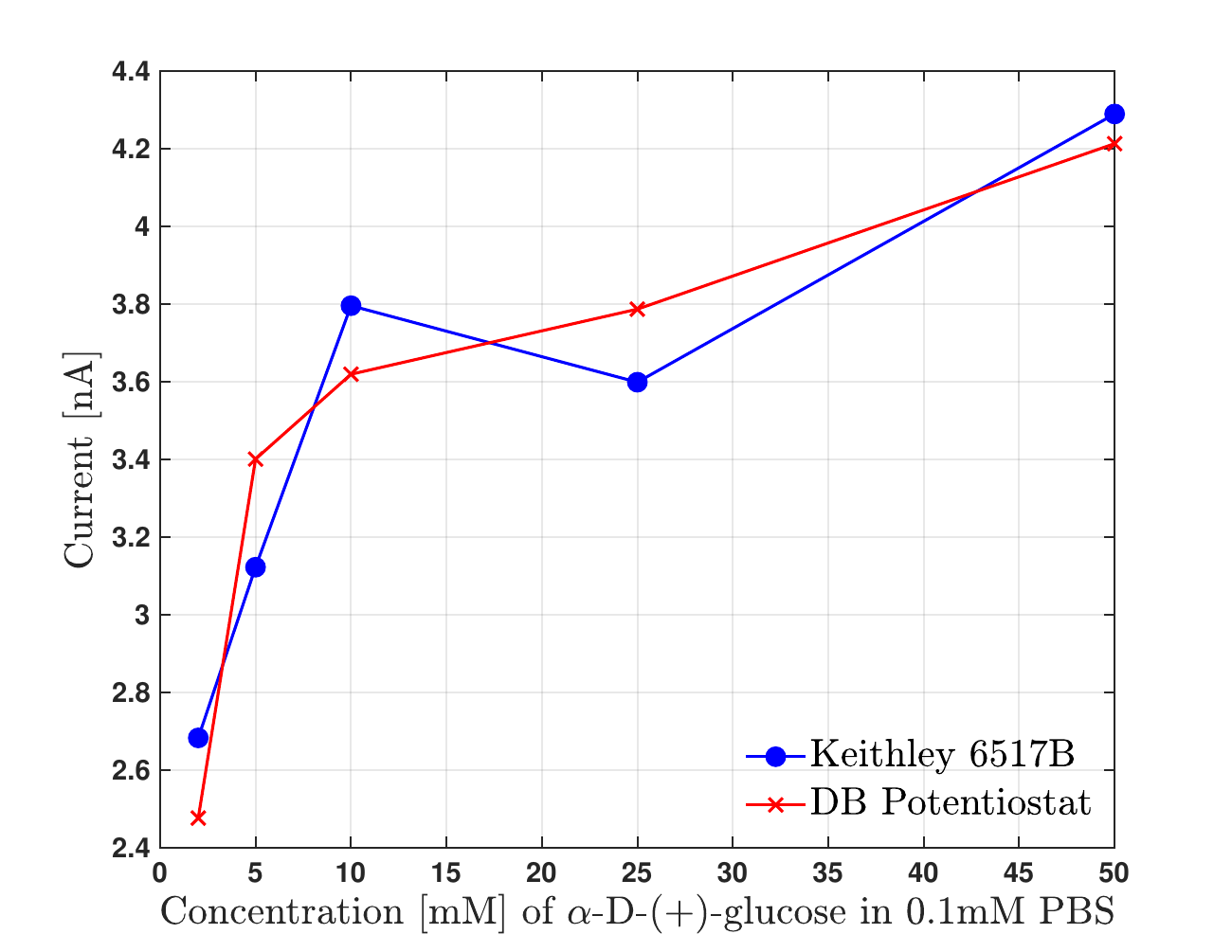} 
        \caption{DB potentiostat calibration curve for non-enzymatic glucose detection.}
   \label{Fig16_DIGOTAvsPGSTAT_glucosio}
\vspace{-20pt}
\end{figure}

\begin{figure}[t]  
    \centering
    \includegraphics[width=0.5\textwidth]{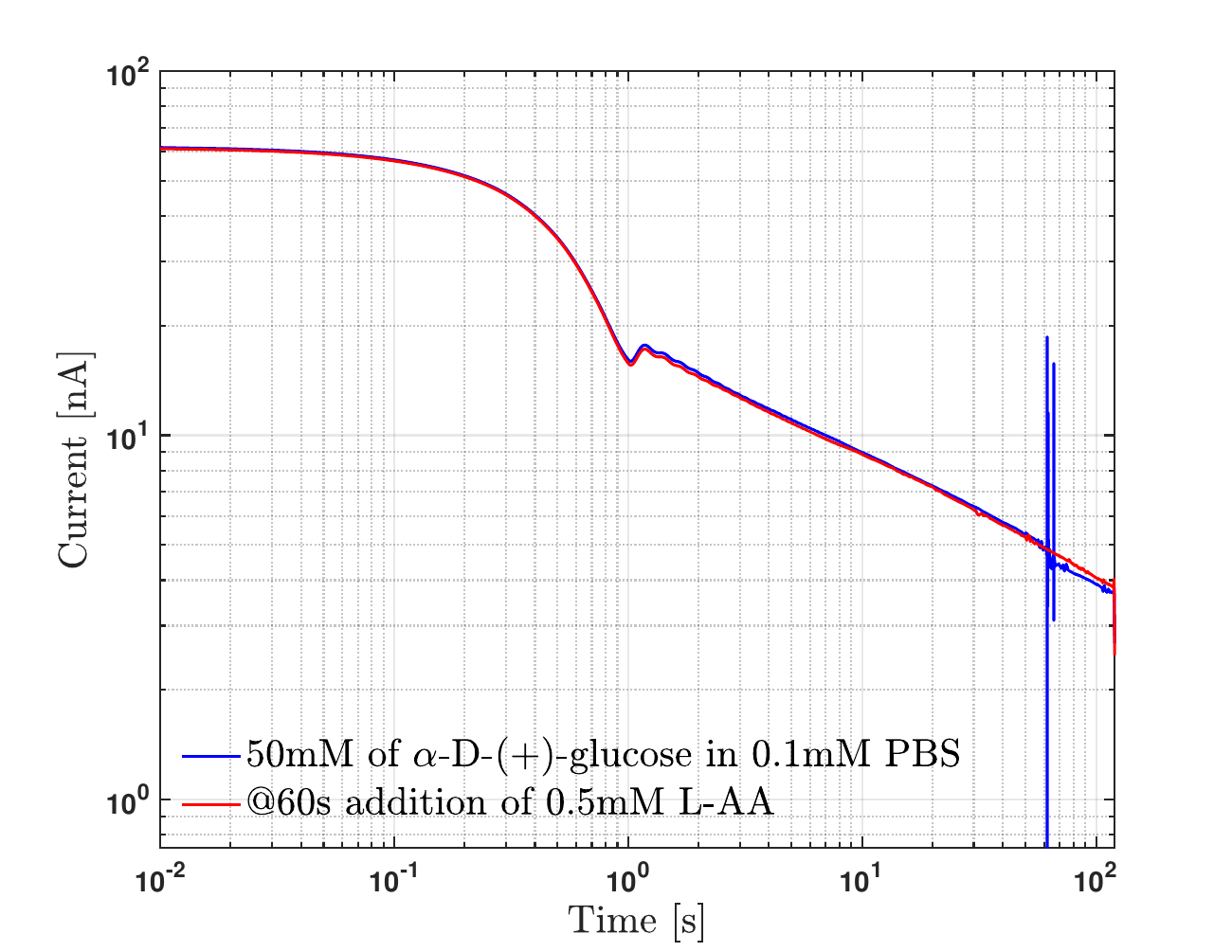} 
        \caption{Current vs time curves showing the responses of H1-ePt to 50\,mM of glucose steady (blue) and (red) during the addition at the time 100\,s of 0.5\,mM of ascorbic acid (L-AA). The spikes in the current correspond to the addition of the interfering solution to achieve a whole concentration of 0.5\,mM of L-AA in 50\,mM of $\alpha$-D(+)-glucose in 0.1\,M PBS aerated.}
   \label{Fig17_CronoGlucosio_LAA}
\vspace{-15pt}
\end{figure}

In practical PoC applications, solution degassing is typically impractical due to its power and time requirements. Therefore, no degassing is performed in this work, and its effects are explicitly considered.


In cyclic voltammetry measurements using bare Pt electrodes of 50\,\textmu m of diameter, reactions of 20\,mM of glucose in aerated 0.1\,M PBS were not visible at all. With the corresponding H1-ePt nanostructure, the voltammetry changes completely and several peaks associated to electrochemical oxidation reactions appear, as seen in Fig.\ref{Fig15a_CVnanostructure}. The electrochemical oxidation of glucose (GOR) on platinum in neutral phosphate buffer proceeds primarily through dehydrogenation at the C1 carbon. 

Among the anomeric forms of glucose, the $\alpha$-glucose employed in this study it the one that exhibits the lowest reactivity, whereas $\beta$-glucose is the most reactive, likely due to the orientation of the anomeric C–H bond. The overall oxidation process is typically described within three distinct potential regions. 
First (region (1)), the “hydrogen region” (from 0 to 350\,mV vs. RHE), in which the platinum uniquely enables a characteristic oxidative response not observed for other electrode materials or for related organic substrates such as alcohols, aldehydes, or acids. Here, glucose oxidation is strongly influenced by the adsorbed hydrogen, and the hemiacetal moiety serves as the principal reactive site. 

Surface-adsorbed glucose or its intermediates govern the reaction pathway. Secondarily (region (2)), the “double-layer region” (from 400\,mV to 800\,mV vs. RHE) exhibits two oxidation peaks at low scan rates. The peak currents diminish as anions or organic specie, in particular glucono-$\delta$-lactone, adsorb onto the surface and it depends on glucose concentration. In the region (1) and (2) the presence of oxygen dissolved in the solution generates an additional reaction, the oxygen reduction reaction (ORR) that compete with the GOR generating a sum of the two current contributions. The only one region in which there is a peak attributed to the complex reaction mechanism of the oxidation of glucose and that has no additional current dependent on mass transport of other species naturally present in the solution is (3), the “oxide region” ($>$ 900\,mV vs. RHE) in which the glucose reacts with platinum oxide, and lactone-type poisoning products undergo further oxidative decomposition.

\begin{table*}[t]
    \centering
    \caption{Potentiostats State-of-Art performance comparison}
    \resizebox{\textwidth}{!}{
{\renewcommand\arraystretch{1} 
    \begin{tabular}{|c|c|c|c|c|c|c|c|c|c|c|c|}
        \hline
        & Units & \cite{21_5_alMamun}&  \cite{24_8_Chilukuri}  & \cite{29_13_Ahmad}& \cite{35_19_chou} & \cite{36_20_rubino} & \cite{37_21_akram} &\cite{41_25_Meimandi} & \multicolumn{3}{|c|}{This Work}\\
        \hline
       Topology & - & I-V (R-TIA) & (C-TIA) & CC & I-F & DB & Digital & DB & \multicolumn{3}{|c|}{DB}\\
       \hline
       Sensing Method & CV, CA & CA & CV, CA & CV, CA & CA & CA & CA & CA & \multicolumn{3}{|c|}{CA}\\
       \hline
       Technology & nm & 180 & 180 & 180 & 180 & 180 & 180 & 180 & \multicolumn{3}{|c|}{130}\\
       \hline
       Supply Voltage & V & 1.8 & 1.8 & 1.5 & 1.2 & 0.4 & 1 - 1.8 & 0.4 & 0.3 & 0.4 & 0.5 \\
        \hline
        cal (calN = calP)& - & - & - & - & - & - & - & - & \multicolumn{3}{|c|}{1}\\
        \hline
        clk Frequency& kHz & - & 5000 & - & - & 50 & 0.1 - 10000 & 25 & \multicolumn{3}{|c|}{50}\\
        \hline
        Power & nW & 22000 & 86400 & 2330000  & 1400 & 4.7 & 3.72 & 1.2 & 1.65 & 2.96 & 53.5  \\
        \hline
        Area &  mm\textsuperscript{2} & 0.0179 & 1.47 & 0.0684 & 0.037 & 0.00046 & 0.165 & 0.00026 & \multicolumn{3}{|c|}{0.011532}\\
        \hline
        Sensing Current Range & A & $\pm$ 10\,\textmu & 1\,n - 10\,\textmu & 4.2\,\textmu - 180\,\textmu & $\pm$2\,\textmu & -22\,n - 33\,n & 80\,p - 240\,\textmu & -12\,n - 24\,n & \multicolumn{3}{|c|}{600\,p - 50\,\textmu}\\
        \hline
        Dynamic Range  & - & 155.92 & 131.4 & - & - & 58 & 129.5 & 60  & 53.82 & 71.42 & 63.20 \\
        \hline
        Linearity & - & 0.999 & - & 0.9 & 0.997 & 0.999 & 0.998 & 0.991 & \multicolumn{3}{|c|}{0.991}\\
        \hline
        Sensitivity  & LSB/mM & - & - & - & - & 5.2 & - & 18,740  & 334,137 & 16,304 & 15,405 \\
        \hline
        Measurements & - & Yes & Yes & Yes & Yes & No & Yes & Yes & \multicolumn{3}{|c|}{Yes}\\
        \hline
        Targeted Analyte & -& Glucose (e) & Dopamine & Glucose (e) & pH, Glucose (e) & Glucose (ne)  & Glucose (e) & Dopamine & \multicolumn{3}{|c|}{Glucose (ne) }\\
        \hline       
    \end{tabular}}
}
\label{SoA table}
    \vspace{-20pt}

\end{table*}

Therefore, among all the peaks promoted by the oxidation of glucose, the only one which does not compete with the ORR happens in the H1-ePt during the first stages of the platinum oxidation at +250\,mV vs Pt (+450\,mV vs Ag/AgCl; +1150\,mV vs RHE) as shown clearly in Fig.\ref{Fig15b_LSVglucose}.

\subsubsection{DB Potentiostat Measurements}
Chronoamperometric measurements with the DB potentiostat were performed in five solutions exploiting the same measurement setup in Fig.\ref{Fig9_MeasurementSetup}, keeping the $f$\textsubscript{clk} = 50\,kHz while the static FI-DIGOTA currents have been set equal to $i$\textsubscript{p} = 4.89\,nA and $i$\textsubscript{n} = 10.16\,nA.
It has been found that H1-ePt surface responded to glucose with two linear dependencies in the glucose concentration range from 2\,mM to 50\,mM with a limit of detection (LoD) of 530\,\textmu M that corresponds to a glucose concentration of 530\,mg/dL in a fluid and a sensitivity of 0,3\,nA mM\textsuperscript{-1}, as demonstrated in Fig.\ref{Fig16_DIGOTAvsPGSTAT_glucosio}.


The selectivity of the analyte is tested introducing 5\,mM of L-ascorbic acid (AA) in the measured glucose solution, to simulate typical concentration of the interfering analyte. Fig.\ref{Fig17_CronoGlucosio_LAA} shows the chronoamperometry of the 50\,mM glucose solution with and without L-AA, demonstrating a negligible contribution of the interfering analyte in the overall reaction current thanks to the selectivity of the nanostructurated microelectrode.
This is confirmed by the chronoamperometry tests performed in the presence of the interferent, which measure the same current in both cases. In particular, the measurement performed with the DB potentiostat with and without the interfering analyte presents a relative error of 0.02\%.

\subsection{Potentiostat SoA Comparison}

The comparison with the state of the art for IC potentiostats is presented in Table\ref{SoA table}. 
It is evident that digital potentiostats \cite{36_20_rubino}\cite{37_21_akram}\cite{41_25_Meimandi} drastically reduce the power consumption with respect to standard topology even if the applicability of this strategy is still limited, and CV can not be performed. 

Despite the area overhead due to the trimmable output stage, the overall area is still smaller than other non-DB potentiostats, and the degrees of freedom ($V$\textsubscript{dd}, cal and $f$\textsubscript{clk}) entailed by the presented DB potentiostat allow to sense a wide electrochemical current range maintaining a minimum power consumption and the value of the supply voltage equal or lower than 500\,mV. Furthermore, the clock frequency value can be decreased to further reduce the power consumption at the expense of current resolution.

Moreover, with the introduction of the nanostructured microelectrodes presented in this work, non-enzymatic detection of glucose at nW-range power and sub-500mV supply voltage is demonstrated for the first time.

\section{Conclusions}

In this paper an ULV, ULP chronoamperometric electrochemical sensor based on a configurable DB-potentiostat in 130nm has been designed and experimentally characterized, revealing the detection of a wide electrochemical current range, spanning from 600\,pA to 650\,nA, with $R^2$=0.991 linearity at only 1.65\,nW (53.5\,nW) power consumption at $V$\textsubscript{dd} = 300\,mV ($V$\textsubscript{dd} = 500\,mV). The signal transfer and noise
characteristics of the DB-potentiostat have been analytically described in the frequency domain for the first time and validated by simulations and experiments.

Then, a nanostructured mesoporous electrode has been specifically co-designed to operate with the DB-potentiostat in an ULV/ULP sensor for non-enzymatic glucose detection. The sensing platform consisting of the DB potentiostat and the electrode has been first validated both under diffusion-controlled regime and then under kinetic-controlled regime for the target glucose monitoring application, demonstrating, for the first time, successful non-enzymatic glucose detection at physiological levels at the lowest reported power (1.65 nW) and at the lowest reported supply voltage (0.3 V), thus revealing a strong potential for emerging PoC diagnostics applications.

\bibliography{bibliography} 
\bibliographystyle{ieeetr}

\end{document}